\documentclass[fleqn,usenatbib]{mnras}
\usepackage{newtxtext,newtxmath}
\usepackage[dvipsnames]{xcolor}

\usepackage[T1]{fontenc}
\usepackage{ae,aecompl}

\usepackage{graphicx}	
\usepackage{amsmath}	
\usepackage{amssymb}	
\usepackage{mathrsfs}
\newcommand{\be}{\begin{equation}}
\newcommand{\ee}{\end{equation}}

\newcommand{\nm}{\log \mbone}
\newcommand{\nq}{\log q}
\newcommand{\gw}{\mathrm{GW}}
\newcommand{\df}{\mathrm{d}}
\newcommand{\bh}{\mathrm{bh}}
\newcommand{\cm}{\mathrm{cm}}
\newcommand{\pc}{\mathrm{pc}}
\newcommand{\mpc}{\mathrm{Mpc}}
\newcommand{\au}{\mathrm{au}}
\newcommand{\yr}{\mathrm{yr}}
\newcommand{\Myr}{\mathrm{Myr}}

\newcommand{\mbhb}{\mathrm{\bullet}}
\newcommand{\erd}{\mathrm{erd}}
\newcommand{\act}{\mathrm{act}}
\newcommand{\ejct}{\mathrm{ej}}
\newcommand{\hvb}{\mathrm{hvb}}
\newcommand{\collision}{\mathrm{quick}}
\newcommand{\chirp}{\mathrm{chirp}}

\newcommand{\bi}{\mathrm{bi}}

\newcommand{\omerger}{\mathrm{eject}}
\newcommand{\imerger}{\mathrm{slow}}

\newcommand{\msone}{m_{\star 1}}
\newcommand{\mstwo}{m_{\star 2}}
\newcommand{\mbone}{M_1}
\newcommand{\mbtwo}{M_2}
\newcommand{\astar}{a_{\star}}
\newcommand{\estar}{e_{\star}}
\newcommand{\acm}{a_{\cm}}
\newcommand{\ecm}{e_{\cm}}
\newcommand{\abh}{a_{\bh}}
\newcommand{\ebh}{e_{\bh}}
\newcommand{\solarm}{M_\odot}
\newcommand{\hubble}{\mathrm{hubble}}

\newcommand*\unit[1]{\mathrm{#1}}
\newcommand*\tab[1]{Table~\ref{#1}}
\newcommand*\fig[1]{Figure~\ref{#1}}
\newcommand*\eqn[1]{Equation~\ref{#1}}
\newcommand*\sect[1]{Section~\ref{#1}}

\title[]{The cosmological distribution of compact object mergers from dynamical interactions with SMBH binaries}

\author[]{
Yi-Han Wang$^{1}$\thanks{rosalba.perna@stonybrook.edu;\newline\qquad yihan.wang.1@stonybrook.edu},
Nathan Leigh$^{1,2,3}$, Alberto Sesana$^{4}$, Rosalba Perna$^{1,5}$
\\
$^{1}$ Department of Physics and Astronomy, Stony Brook University, Stony Brook, NY 11794-3800, USA\\
$^2${Departamento de Astronom\'ia, Facultad de Ciencias F\'isicas y Matem\'aticas,
Universidad de Concepci\'on, Concepci\'on, Chile}
\\
$^3$Department of Astrophysics, American Museum of Natural History, Central Park West and 79th Street, New York, NY 10024
\\
$^{4}$ School of Physics and Astronomy and Institute of Gravitational Wave Astronomy, University of Birmingham, Ed\
gbaston B15 2TT, UK\\
$^{5}$
{Center for Computational Astrophysics, Flatiron Institute, 162 5th Avenue, New York, NY 10010, \
USA}
}

\pubyear{2019}
\begin{document}
\label{firstpage}
\pagerange{\pageref{firstpage}--\pageref{lastpage}}
\maketitle


\begin{abstract}
    We combine sophisticated high precision scattering experiments, together with results from the Millenium-II simulation,  to compute the cosmic merger rate of bound compact object (CO) binaries dynamically interacting with supermassive black hole binaries (SMBHBs). We consider binaries composed of white dwarfs (WDs), neutron stars (NSs) and black holes (BHs).
    The overall merger rates for WD-WD, NS-NS, BH-BH, BH-NS binaries and EBBH (eccentric binaries of black holes) from redshift $\sim 5$ are found to be $4.32\times 10^3\,\yr^{-1}$($5.93\times10^2\,\yr^{-1}$ for Type Ia SNe), $82.7\,\yr^{-1}$, $96.3\,\yr^{-1}$, $13.1\,\yr^{-1}$ and $148\,\yr^{-1}$ , respectively, {for a nominal CO binary fraction in the Galactic centre of 0.1.} We calculate the distance ($R$) distribution of the merger sites with respect to the host galaxies of the binaries. The distribution shows a wide range of distances  up to $\sim \mpc$; this tail is produced by escaped hyper-velocity CO binaries. Due to the differences in the matter density of the surrounding environment, merger events with different $R$ are expected to display significantly different signatures in their EM counterparts.  In particular, merger events (and especially NS-NS) producing a relativistic jet but occurring in the intergalactic medium will have very weak afterglow radiation relative to their prompt emission. These events, which we call 'off-center', can only be produced from a close encounter between  CO binaries and SMBHBs; hence the detection of such merger events would indicate the existence of nearby SMBHBs, and in particular with high mass ratio, produced in the aftermath of a major galaxy merger. 
\end{abstract}

\begin{keywords}
Galaxy: centre -- Galaxy: kinematics and dynamics -- stars: black holes -- stars: kinematics and dynamics -- galaxies: star clusters: general -- stars: neutron
\end{keywords}



\section{Introduction}\label{sec:intro}

The detection of binary black hole (BH) and binary neutron star (NS) mergers via gravitational waves (GWs)\citep[e.g.,][]{2019arXiv190201557T,2016PhRvL.116f1102A,2018arXiv181112940T,2017PhRvL.119p1101A} has opened a new window to explore the Universe. A lot of interest has been devoted to understanding the physical origins of these events, and in particular the formation scenarios of the binaries which merge. The most popular hypotheses 
involve field binaries \citep[e.g.,][]{2003MNRAS.341..385P, 2013ApJ...779...72D, 2014ApJ...789..120B, 2016Natur.534..512B, 2016A&A...588A..50M, 2016MNRAS.458.2634M} and dynamical formation \citep[e.g.,][]{2000ApJ...528L..17P,2006ApJ...637..937O, 2009ApJ...692..917M, 2013MNRAS.429.2298M, 2014MNRAS.444...29L, 2016ApJ...824L...8R, 2018ApJ...866L...5R, 2016ApJ...816...65A, 2017ApJ...834...68C, 2018PhRvD..97j3014S, 2018MNRAS.481.5445S, 2018MNRAS.478.4030G, 2019MNRAS.486.5008A, 2018MNRAS.481.5123B, 2018MNRAS.473..909B, 2018PhRvL.121p1103F, 2018arXiv181110627F, 2019arXiv190100863D,Perna2019}.

Once binaries are formed, their subsequent evolution up to merger, and hence the locations of their mergers, are also of great importance. Compact object (CO) binaries located in the vicinity of a supermassive black hole (SMBH) are subject to Lidov-Kozai (LK) oscillations, which accelerate their merger times, thus enhancing their merger rates \citep[e.g.,][]{2017ApJ...841...77A, 2017ApJ...836L..26C, 2016ApJ...831..187A, 2012ApJ...757...27A, 2002ApJ...578..775B, 2018ApJ...856..140H, 2017ApJ...836...39S, 2016ApJ...828...77V, 2016ARA&A..54..441N, 2003ApJ...598..419W}. This channel of CO mergers has been shown to have rates which are competitive with other proposed channels, due to the combination of the large star density in the vicinity of the central SMBH, and the shorter merger times resulting from dynamical interactions.

The phenomenology associated with dynamical interactions becomes richer when two SMBHs are paired together, making up an SMBH binary (SMBHB). In this case, it has been shown that, in addition to Hypervelocity stars (or CO), also Hypervelocity binaries (HVBs) can be produced by the dynamical interaction between the CO binary and the SMBHB \citep[][]{2007ApJ...666L..89L,2009MNRAS.392L..31S,2018MNRAS.475.4595W,2019MNRAS.482.3206W,Coughlin2018}. Therefore, detection of HVBs can serve as an unique diagnostic for the presence of a secondary SMBH in galactic nuclei. 

Besides ejection, another outcome of strong dynamics around SMBHs is a stellar tidal disruption event \citep[TDE, e.g.,][]{1988Natur.333..523R}. TDEs can also be used to probe SMBHBs in many different ways \citep[e.g.,][]{2017MNRAS.469.2042B,2017MNRAS.471L.115C,2018MNRAS.477.4009D}.

TD rates have been found to be particularly enhanced by unequal SMBHBs, with mass ratios $<0.1$ \citep{2011ApJ...729...13C,2018arXiv181101960T}. Moreover, since main sequence stars are generally disrupted only by SMBHs with masses below $10^8\solarm$ (more massive black holes just swallow the whole star without tearing it apart), a TDE flare in a galaxy expected to host a billion solar mass SMBH would be indicative of the presence of a secondary, much lighter, SMBH companion, as proposed by \citet{2018MNRAS.479.3181F}. Interestingly, few peculiar transient events have been interpreted as candidate TDEs from low mass ratio SMBHB binaries \citep{2009ApJ...706L.133L,2018MNRAS.474.3857C}. Conversely, the HVB mechanism quantified by \citet{2018MNRAS.475.4595W} operates most efficiently at SMBHB mass ratios close to unity. Hence, the combination of these two diagnostics cover a significant fraction of the total parameter space available to SMBHBs. The identification of traditional, main sequence star HVBs is limited to the Milky Way and the local group. On the other hand, due to their compactness, the parameter space available to HVBs is much higher for degenerate stars than for main-sequence stars, since binaries containing remnants, especially neutron stars (NSs) and black holes (BHs), can exist at smaller orbital separations. It is therefore interesting to model the generation of CO-HVB ejection from SMBHBs and their potential observable features, which is one of the main focuses of this work.

The observable phenomenology associated with a merger event in the electromagnetic (EM) domain depends on the type of objects making up the CO binary.
Mergers of two NSs have been confirmed to be associated with short gamma ray bursts (GRBs) \citep{Abbott2017}, and it is expected that this should also be the case for mergers of NS-BH binaries for small mass ratios (see e.g. the review in \citealt{Bartos2013}). BH-BH mergers are most likely to be EM-quiet, albeit ideas for accompanying radiation have been proposed \citep{Perna2016,Loeb2016,Zhang2016,Woosley2016,DeMink2017,Bartos2017,Liebling2016,Murase2016,Janiuk2017,Fraschetti2018}.
The location of the merger site has long been considered as an important diagnostic of the progenitor type. For CO mergers, the predictions are of a wide distribution of distances from the galaxy hosts, primarily reflecting the natal kicks, the merger times, and the mass of the host galaxy (e.g. \citealt{Belczynski2006,Oshau2017,Perna2018}). For all three types of CO binaries (i.e. NS-NS, NS-BH and BH-BH), a small fraction of events at $\sim$~Mpc scale is expected only in small galaxies, with masses $\lesssim 10^{10}M_\odot$ 
(e.g. \citealt{Perna2002}), and hence yielding a small contribution to the total merger rates.
For larger, Milky-Way type of galaxies, the gravitational potential of the galaxy prevents very large distances to be reached before mergers. Therefore, the production of HVBs presents the special possibility that some fraction of short GRBs may occur in isolated regions of the Universe between galaxies, or in the intergalactic medium (IGM), and not be directly associated with any particular host galaxy.  These isolated short GRBs could thus indicate the presence of an SMBH-IMBH/SMBH binary hosted in the nearest galaxy or galaxy group/cluster.

Identifying these off-host GRB events as a function of redshift opens up the possibility of using them to constrain the SMBHB fraction in galaxies and how it changes as a function of cosmic time.  The sophisticated $N$-body simulations performed by \citet{2019MNRAS.482.3206W} revealed how the post-ejection properties of the HVBs depend on the properties of the ejecting SMBHB.  Thus, combining observations of the electromagnetic properties of off-host GRBs with gravitational wave detection -- specifically the information extracted from the waveform data -- could further constrain the properties of the SMBHBs as well as the environment in their immediate vicinity. Moreover, if Type Ia supernovae result from mergers of two WDs \citep[i.e.,][]{1984ApJS...54..335I}, the detection of 'orphan' (i.e. in the IGM, away from their host galaxies) Type Ia SNe, could provide another powerful probe of SMBH binarity. We will in fact see that hypervelocity WD-WD binaries are also expected to be produced by dynamical interactions with SMBHBs.

In this paper, we combine the cosmic SMBHB population extracted from a large scale cosmological simulation \citep[The Millennium-II,][]{2009MNRAS.398.1150B} with high-resolution, N-body simulations of the interaction of CO binaries with those SMBHBs, to predict the cosmological distribution of CO binaries, and their properties (including radial separation from their host galaxies) at the time of merger.
Additionally, we compute the cosmological merger rate of these events, and their contribution to the GW signal, and discuss our results in the context of present and future GW observations and accompanying electromagnetic counterparts.

 The paper is organized as follows.  In \sect{sec:SMBHBformation}, we discuss the cosmic SMBHB formation rate and cusp erosion process that provides the ideal environment to produce various types of CO merger events. In \sect{sec:relfraction} we perform the scattering experiments in the reference frame of the SMBHB to obtain the relative fraction of different types of merger events and their dependences on initial parameters, and discuss the properties of the outcomes. In \sect{sec:cosmic}, we combine the outcomes of the scattering experiments and SMBHB formation rate to estimate the cosmic rates of different types of merger events triggered by the SMBHBs.  Our conclusions are summarized in \sect{sec:summary}. 

\section{SMBHB formation and cusp erosion}\label{sec:SMBHBformation}

\begin{figure*}
 \includegraphics[width=\textwidth]{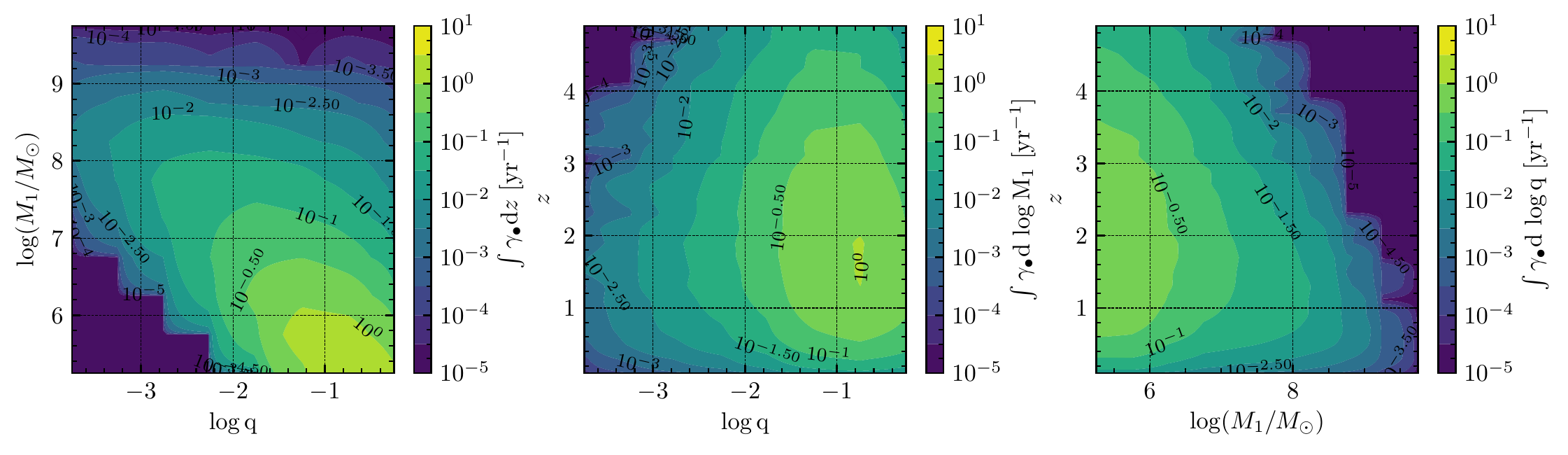}
 \caption{ The observed SMBHB formation rate derived from the Millennium II simulation as a function of the primary SMBH mass $\mbone$, the binary mass ratio $q$, the redshift $z$ and the binary semi-major axis $a_\bh$. These distributions are used in our simulations to sample the parameters of the SMBHBs.}
 \label{fig:GammaSMBHB}
\end{figure*}

\subsection{SMBHB formation rate}
{ The cosmic population of SMBHBs is derived using the data from Millennium-II \citep[e.g.,][]{2009MNRAS.398.1150B}, which is a large scale N-body simulation with sufficient resolution to distinguish dark matter halos down to $\sim10^8\solarm$. On top of this dark matter backbone, semianalytic galaxy evolution models have been included to reconstruct the cosmic history of galaxies down to $\approx10^6\solarm$, hosting central black holes with masses down to $\sim 10^4M_\odot$. In particular, data are taken from the semianalytic model developed in \citet[][]{2011MNRAS.413..101G}, following the procedure described in \citet[][]{2018arXiv181101960T}. For consistency, all the post processing is performed using the same cosmology as in Millennium-II: $h=H_0/(100\,\mathrm{km\,s^{-1}\,Mpc^{-1}})=0.73$, $\Omega_\Lambda=0.75$ and $\Omega_M=0.25$. Although these values of the cosmological parameters are outdated, a $\sim$ few percent level change in $h$, $\Omega_\Lambda$ or $\Omega_M$ does not have a significant impact on our results.}

We select all mergers involving two galaxies both hosting an SMBH, for a total of $N_\bullet=169435$ events across the history of the Millennium-II simulation. We then construct a 3D grid by logarithmically binning $\log_{10}(\mbone)$ and the mass ratio $q$ within the ranges $\mbone\in[10^5,10^{10}]M_\odot$ and $q\in[10^{-4},1]$, and by evenly binning the redshift $z$ from 0 to 5 according to the redshift of each Millennium-II snapshot. The redshift of each merger is identified with the redshift at which the progenitor galaxy is identified in the simulation. In doing this, we are implicitly assuming that the timescale for dynamical friction (DF) to bring the two SMBHs together in a relatively close binary following the galaxy merger is shorter than the typical time elapsed between two snapshots of the simulation, which is $\approx 300$ Myr. Although the DF timescales might be longer \citep{1987gady.book.....B}, we ignore this complication in this work.

Dividing the SMBHB number in each bin by the co-moving volume $V_c$ of the Millennium-II simulation, we get an approximation of the differential SMBHB formation rate, 
\begin{equation*}
\frac{\df^4 N_\bullet}{\df z\,\df\nq\,\df\nm\,\df V_c}\,.
\label{eq:SMBHBdens}
\end{equation*}
 This quantity is simply the differential number density of SMBHBs per unit redshift, logarithmic mass and mass ratio. If we now multiply it by the standard comoving volume shell ${\df V_c}/{\df z}$ and convert redshift into time by multiplying ${\df z}/{\df t_r}\times{\df t_r}/{\df t}$ with $t_r=t/(1+z)$, we finally obtain the differential SMBHB formation rate observed on Earth,
\be
\gamma_\mbhb(z,\nq,\nm) = \frac{\df^4 N_\bullet}{\df z\,\df\nq\,\df\nm\,\df t}\,.
\ee

\fig{fig:GammaSMBHB} shows the distribution of the SMBHB formation rate $\gamma_\mbhb(z,\nq,\nm)$ in a 2-D plane by integrating over each single variable $z$, $\nm$ and $\nq$ separately. 
The SMBHB formation rate as observed on Earth is dominated by systems with $M<10^6$M$_\odot$, with mass ratios broadly distributed around $0.1$, reaching out to $z\approx 4$.

\subsection{The SMBHB-bound binary interaction rate}
\label{sec:intrate}

If we take a snapshot of the sky, the differential number of SMBHBs that are actively interacting with the surrounding population of bound stars is given by the differential SMBHB formation rate $\gamma_\mbhb(z,\nq,\nm)$ times the duty cycle $D_\erd(z,\nq,\nm)$ that the SMBHB takes to erode the cusp of stars bound to $\mbone$. Note that $D_\erd$ is generally a function of $z,\nq,\nm$. During this duty cycle, the SMBHB interacts with a mass $M_\erd(z,\nq,\nm)$, so that the rate at which the SMBHB interacts with bound stars is simply $\dot{M}_{\erd}=M_\erd/D_\erd$ (where we omitted the parametric dependencies). The differential interaction rate is therefore given by the differential number of active SMBHBs times the rate at which those binaries interact with ambient stars:
\begin{eqnarray}
\dot{\mathcal{M}}_\act(z,\nq,\nm)&=&\frac{\df^4 M_\act}{\df z\,\df\nq\,\df\nm\,\df t}\\
&=&\gamma_\mbhb D_{\erd} \dot{M}_{\erd}\\
&=& \frac{\df^4 N_\bullet}{\df z\,\df\nq\,\df\nm\,\df t}{M}_{\erd}\,.
\label{eq:Mactive}
\end{eqnarray}
Therefore, to know the differential interaction rate, we simply need to estimate the total mass interacting with the binary, ${M}_{\erd}$. Numerical simulations of unequal mass SMBHBs embedded in stellar cusps have shown that the secondary hole efficiently inspirals down to a separation at which the mass in stars enclosed within its orbit around $\mbone$ is of the order of $\approx 2\mbtwo$, and it does this without significantly affecting the stellar distribution \citep[e.g.,][]{2007ApJ...656..879M}. At that point, further shrinking of the orbit proceeds via efficient ejection of bound stars, thus eroding the cusp \citep[e.g.,][]{2008ApJ...686..432S}. Pending numerical factors of order unity, the number of interacting stars can therefore be approximated as $M_\erd\sim\mbtwo=q\mbone$.

The specific { mass interaction rate} that is observed on Earth up to redshift $z$ can then be computed by integrating $\gamma_\erd(z,\nq,\nm)$ from 0 to $z$,
\be
\Lambda(z,\nq,\nm) = \int_0^z\dot{\mathcal{M}}_\act(z^\prime,\nq,\nm) \df z^\prime\,.
\ee
\begin{figure*}
 \includegraphics[width=\textwidth]{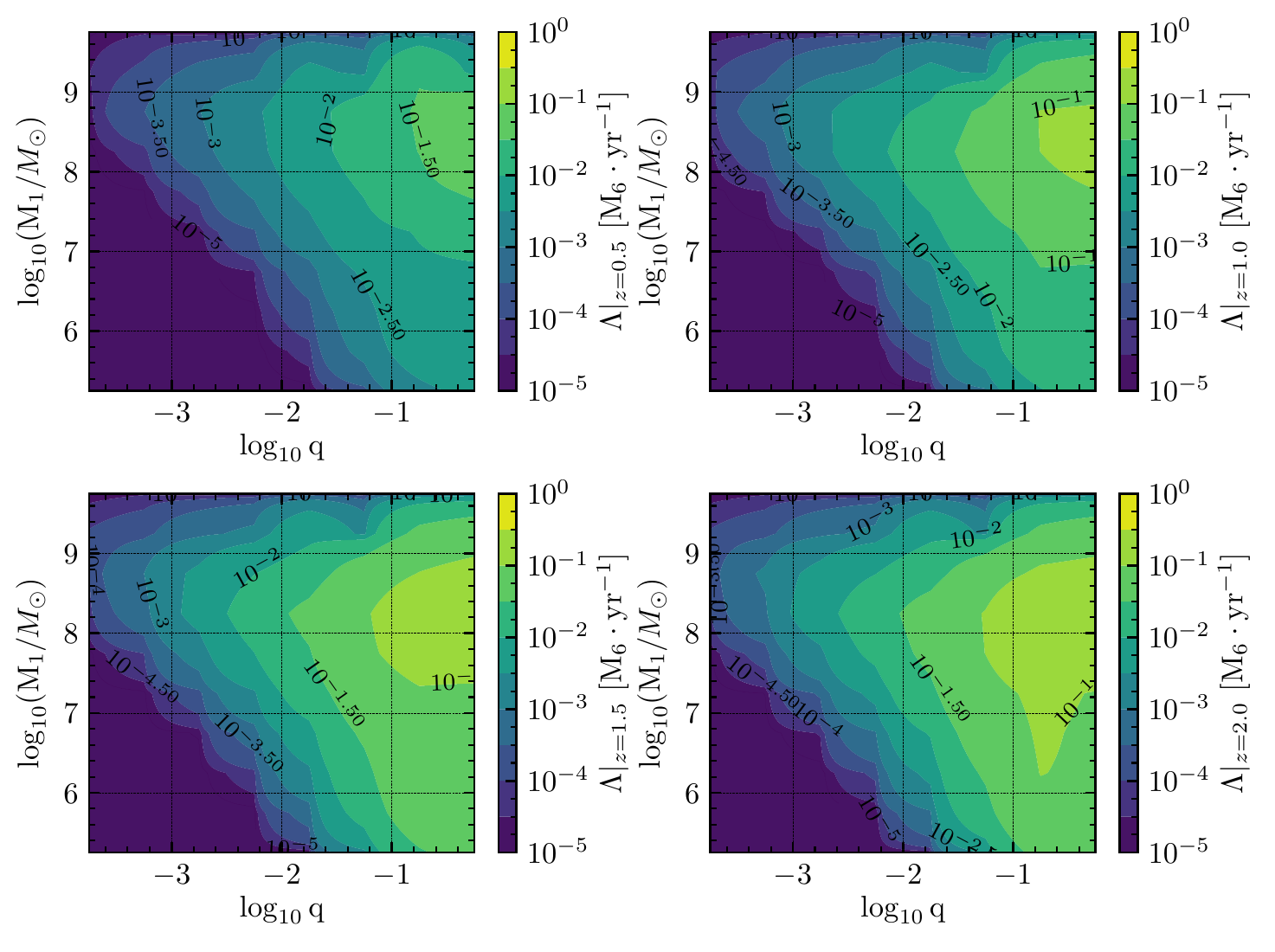}
 \caption{Contour plots of the mass in CO binaries that is actively interacting with a SMBHB as a function of SMBHB primary mass and mass ratio. The four panels represent this quantity integrated up to different redshifts $z=0.5$, $z=1$, $z=1.5$ and $z=2$.}
 \label{fig:most_interact}
\end{figure*}

\fig{fig:most_interact} shows $\Lambda(z,\nq,\nm)$ at different redshifts: $z=0.5$, $z=1.0$, $z=1.5$ and $z=2.0$. This figure indicates that for relatively low redshifts, most of the CO binaries that are actively interacting with SMBHBs correspond to high mass ratio (0.1-1) SMBHBs with primary masses $\mbone$ between $10^6\,M_\odot$ and $10^9\,M_\odot$. { Integrating the contours in the figure yields that about $10^6\,M_\odot$ of stars and COs actively interacting with SMBHBs out to $z=2$ is potentially observable from Earth every year.}

In order to set up the scattering experiments in \sect{sec:relfraction}, we need an estimate of the SMBHB separation at which stars are efficiently ejected from the cusp. This corresponds to $a_0$ so that $M_*(r<a_0)=2\mbtwo$. For simplicity, following \citet[][]{2010ApJ...719..851S}, we assume the stellar density profile to follow an isothermal sphere. With this assumption, and by using the $M-\sigma$ relation from \citet[][]{2000ApJ...539L...9F}, we get 
\be\label{eq:a0}
a_0\sim 0.8\bigg(\frac{q}{1+q}\bigg)\bigg(\frac{\mbone}{M_6}\bigg)^{1/2}~\mathrm{pc}\,.
\ee

\section{Scattering experiments in the reference frame of the SMBHB}\label{sec:relfraction}

Not all actively interacting CO binaries will produce outcomes of interest to us, in particular hypervelocity binaries that escape their host galaxies and merge in the intergalactic medium. Therefore, scattering experiments between different types of CO binaries with SMBHBs with different $\mbone$ and $q$ values are needed to obtain good statistics in the relevant regions of parameter space.

\subsection{Numerical setup}

We perform extensive scattering experiments using our high-precision code {SpaceHub} (Yihan 2019 in prep.; see also Yihan et al. 2019 for more details), which employs the {\tt ARCHAIN} algorithm \cite[e.g.,][]{1993CeMDA..57..439M} to accurately trace the motion of tight binaries with arbitrarily large mass ratios and eccentricities. The original chain structure in the algorithm, combined with the Kahan summation introduced by us to the regularized code, significantly reduces the round-off errors during close encounters.

\begin{figure}
\center
 \includegraphics[width=0.8\columnwidth]{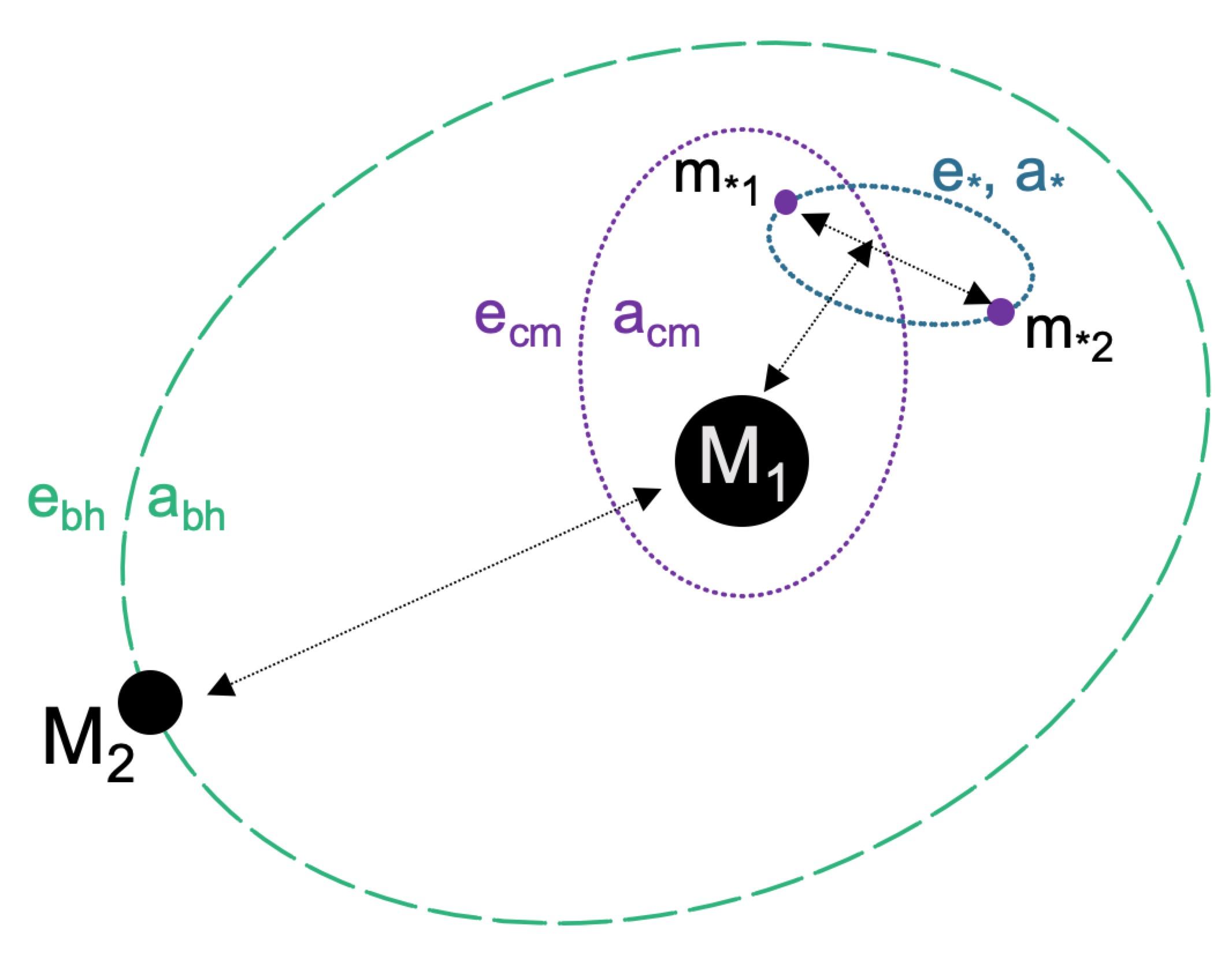}
 \caption{Schematic illustration of the stellar binary orbiting around the super-massive black hole binary (not to scale).}
\label{fig:config}
\end{figure}

In order to cover the relevant SMBHB masses and mass ratios, as indicated by \fig{fig:GammaSMBHB} and \fig{fig:most_interact}, the $q$ and $\mbone$ values of the SMBHBs are logarithmically generated in the range $[10^{-4},1]$ and $[10^5,10^{10}]\,M_\odot$, respectively. The semi-major axis $\abh$ of the SMBHB is set to $a_0/3$, where $a_0$ is the active ejection radius and the eccentricity of the SMBHB $\ebh$ is uniformly distributed in the range $[0,1]$. The choice of  $a_0/3$ is motivated by the fact that N-body simulations and scattering experiments have shown that the whole cusp erosion shrinks the semi-major axis of the SMBHB by roughly a factor of 10  \citep[e.g.,][]{2008ApJ...686..432S,2011MNRAS.415L..35S} starting from $a_0$ given by \eqn{eq:a0}. Therefore, our scattering experiments are performed at a specific semi-major axis of the SMBHB $\abh = a_0/3$, which is the geometric average of $a_0$ and $a_0/10$. \fig{fig:config} shows the configuration of our scattering experiments and corresponding orbital parameters.

We explore five types of stellar mass binaries: WD-WD, BH-BH, NS-NS, BH-NS and EBBH (eccentric binary BHs). {250k scattering experiments are performed for BH-BH, NS-NS and BH-NS binaries while 500k scattering experiments are performed for WD-WD binaries and EBBHs.}

\subsubsection{WD-WD binary population}
For the WD-WD binaries, we take the masses $m_1$ and $m_2$ of each  WD from a normal distribution with $\langle M_{WD}\rangle=0.619 M_\odot$ and $\sigma_m = 0.108 M_\odot$ \citep[e.g.,][]{Tremblay2016}. The separation $a$ of the WD-WD binary is taken from the distribution \citep[e.g.][]{Maoz2017}
\be
N(x)\propto x^3 \ln ( 1 + x^{-4})\,,
\ee
where 
\be
x=\frac{a}{(Kt_0)^{1/4}}
\ee
with 
\be
K = \frac{256}{5}\frac{G^3}{c^5}m_1m_2(m_1+m_2)\,,
\ee
and $t_0$ is the age of the host galaxy. This is evidently a time-dependent model in which high redshift WD-WD binaries would be more compact. However, for simplicity in our simulations, we take $t_0 = 13.6~{\rm Gyr}$ throughout. The eccentricity $e$ of the WD-WD binaries is drawn from a uniform distribution within $[0, 1]$ \citep{Geller2019}.
 The radius of the WD is set to be $0.01R_\odot({m_\odot}/{m_\mathrm{wd}})^{1/3}$.

\subsubsection{BH-BH, NS-NS and BH-NS binary populations}
For the BH-BH, NS-NS and BH-NS binaries, we take the distributions for $\msone,\, \mstwo,\,\astar$ and $\estar$ from the BSE code \href{https://www.syntheticuniverse.org/}{\tt StarTrack} \citep{Belczynski2008} with metallicity $Z=0.1Z_\odot$. 
The BH radii are set equal to their Schwarzschild radii, while the  radii of the NSs are set to be 10 $\mathrm{km}$.

\subsubsection{Eccentric BH-BH binary population}
The properties of the CO binary populations in galactic nuclei are essentially unknown. \citet{Gond2018} suggested that eccentric BH-BH binaries can form via dynamical capture. Therefore, we perform an additional set of simulations with eccentric binary BHs (labelled EBBHs). Following \citet{Gond2018}, the masses of the BHs are drawn from a Salpeter initial mass function (IMF), $N(m) \propto m^{-2.35}$ \citep[e.g.,][]{1955ApJ...121..161S}. The separation is drawn from a logarithmically uniform distribution where $a_{\rm min}=0.01\,\au$ and $a_{\rm max}=1000\,\au$. The circularity $c=1- e^2$ also obeys a logarithmically uniform distribution, where $c_{\rm min} = 10^{-6}$ and $c_{\rm max} = 1$.\\\\

\begin{figure*}
 \includegraphics[width=\textwidth]{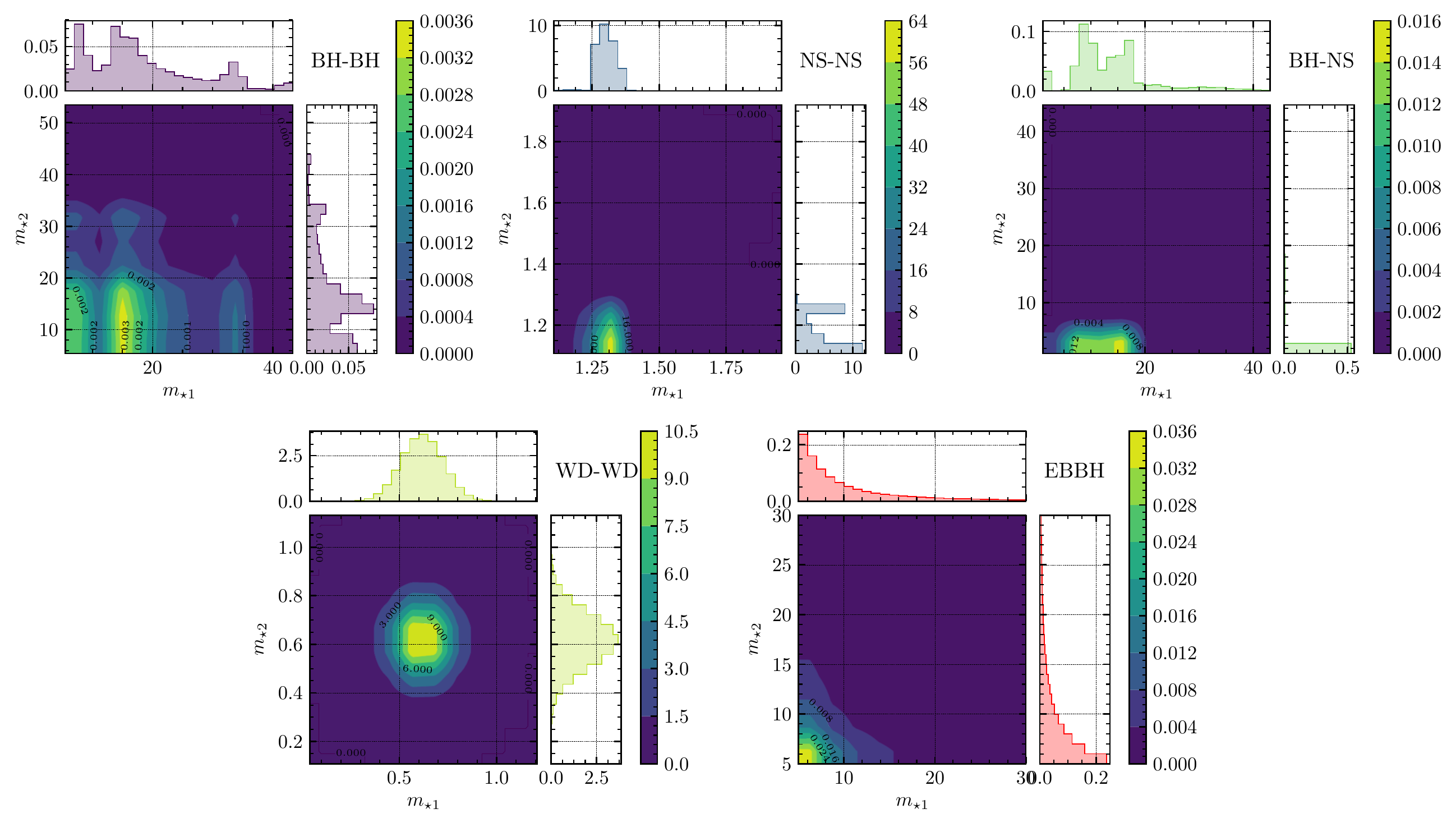}
 \rule{\textwidth}{0.02cm}
 \includegraphics[width=\textwidth]{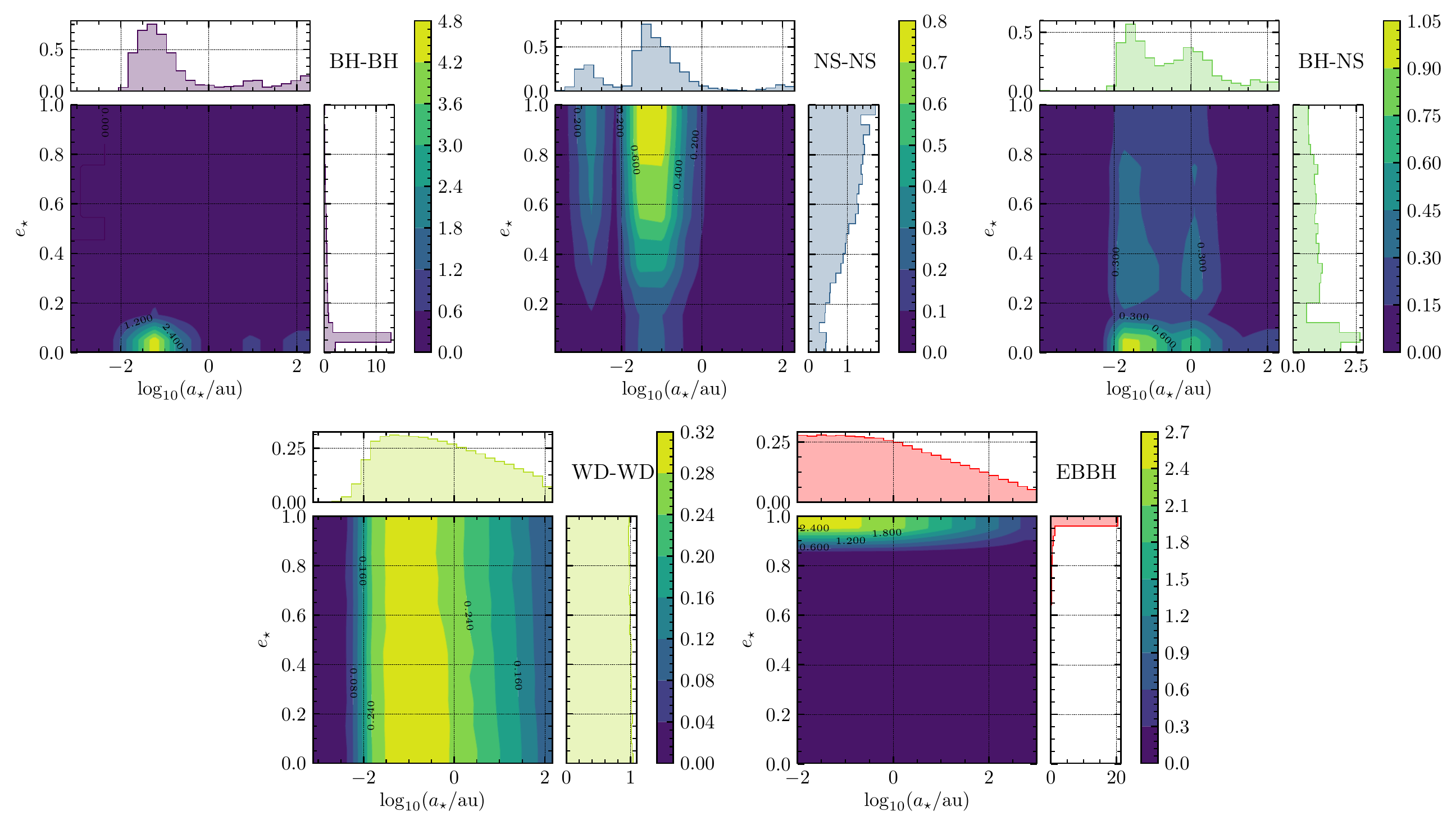}
 \caption{Distribution of the compact object binary population in the $\msone$-$\mstwo$ (top series of panels) and $\astar$-$\estar$ (bottom series of panels) plane. In each panel, the main plot shows the 2D probability density distribution (PDF), whereas the side plots show the 1D distribution marginalized over the other quantity. \textit{Top Left} panel: BH-BH; \textit{Top middle} panel: NS-NS; \textit{Top right} panel: BH-NS; \textit{Bottom left} panel: WD-WD; \textit{Bottom right} panel: EBBH. }
 \label{fig:COb-pop}
\end{figure*}

\fig{fig:COb-pop} shows the parameters of the five types of CO binaries described above. For the BH-BH case, almost all binaries have circular orbits.  Conversely, NS-NS binaries tend to be eccentric, and NS-BH binaries tend to be uniformly distributed in eccentricity. The semi-major axes of the BH-BH, NS-NS and BH-NS binaries are mostly clustered in the $10^{-2}$AU-1AU range, thus forming a population of relatively tight binaries.

 The orbits of the centre of mass (CoM) of the CO binaries are created such that $\acm$ is isothermally distributed (i.e., $p(a)\propto a^{-2}$) around the primary SMBHB in the range of $[R_t, 10\abh]$, where
\be
R_t = 3.7\bigg( \frac{\mbone}{\msone+\mstwo}\bigg)^{1/3}\frac{ 1+\estar}{1-\ecm}\astar
\ee
is the minimum semimajor axis of the stellar binary CoM around $M_1$ ensuring that the stellar binary will not be tidally disrupted at its apocentre when its CoM orbit approaches pericentre around $M_1$.

All angles and inclinations are sampled according to a spherically symmetric distribution. All anomalies are sampled uniformly in time. For each simulation, at least 50 cycles of the CoM orbit of the CO binary are first performed in isolation to ensure that the CO binaries are stable before interacting directly with the SMBHB. 

\subsection{Stability Map}
\label{sec:stability}

Since stars reside deep in the potential well of $\mbone$, relaxation processes due to the extended stellar distribution are negligible and the typical star-SMBHB interaction time is much shorter than the relaxation time $T_{\rm relax}$ of the surrounding stellar system. The cusp erosion can therefore be treated as a set of individual star-SMBHB interactions that can be tackled by means of 3-body scattering experiments. If a fraction of ambient stars is in fact made up by binaries of compact objects (WDs, NSs or BHs) then, during the 4-body interaction, several processes may lead to CO binary mergers.

To understand the different physics behind the merger events, we follow the stability map method in our previous work \citep[][]{2019MNRAS.482.3206W}. Briefly, the four body system can be divided into two triples: an inner triple formed by $\msone$-$\mstwo$-$\mbone$, and an outer triple formed by $\mbone$-$\mbtwo$-$(\msone,\mstwo)$. In each triple, the system can be either stable or unstable. In stable systems, secular effects like LK oscillations may accelerate the merger process. In unstable systems, there is a significant chance of a close interaction occurring between two or more of the components of the triple. The stability of each triple is expressed in terms of the stability factor $S$. For the inner triple we have
\begin{equation}
S_{\mathrm{in}} = \log\bigg(\frac{\acm(1-\ecm)}{\astar(1+\estar)} / Y_\mathrm{in,crit}\bigg)
\end{equation}
where
\begin{equation}\label{eq:crit1}
Y_\mathrm{in,crit} = \frac{3.7}{\beta_\mathrm{out}}-\frac{2.2}{1+\beta_\mathrm{out}} + \frac{1.4}{\beta_\mathrm{in}}\frac{\beta_\mathrm{out}-1}{\beta_\mathrm{out}+1}
\end{equation}
\citep[e.g.,][]{1996ASPC...90..433K} with $\beta_\mathrm{in}=\mathrm{max}\bigg(\frac{\msone}{\mstwo}, \frac{\mstwo}{\msone}\bigg)^{1/3}$, and $\beta_\mathrm{out}=\bigg(\frac{\msone+\mstwo}{\mbone}\bigg)^{1/3}$. 
For the outer triple system we have
\begin{equation}
S_\mathrm{out} = \log\bigg(\frac{\abh(1-\ebh)}{\acm(1+\ecm)} / Y_\mathrm{out,crit} \bigg) 
\end{equation}
where,

\begin{equation}\label{eq:crit2}
Y_\mathrm{out,crit} = \frac{3.7}{\zeta_\mathrm{out}}-\frac{2.2}{1+\zeta_\mathrm{out}} + \frac{1.4}{\zeta_\mathrm{in}}\frac{\zeta_\mathrm{out}-1}{\zeta_\mathrm{out}+1}
\end{equation}
with $\zeta_\mathrm{in}=\bigg(\frac{\mbone}{\msone+\mstwo}\bigg)^{1/3}$, and $\zeta_\mathrm{out}=\bigg(\frac{\msone+\mstwo+\mbone}{\mbtwo}\bigg)^{1/3}$.

If $S_\mathrm{in,out}>0$, then the corresponding triple system is stable, otherwise it is unstable and tends to be disrupted.
We can thus divide the parameter space of the four-body system into four parts, where $(S_\mathrm{in}>0,S_\mathrm{out}>0)$, $(S_\mathrm{in}<0,S_\mathrm{out}>0)$, $(S_\mathrm{in}>0,S_\mathrm{out}<0)$ and $(S_\mathrm{in}<0,S_\mathrm{out}<0)$. For stable triple systems where $S_\mathrm{in,out}>0$, hierarchical LK oscillations can occur. In such systems, the long time scale secular perturbations induced by the outer orbit can drive the eccentricity of the inner orbit to extreme values, while simultaneously decreasing the inclination between the inner and outer orbital planes and conserving angular momentum. 

\subsection{Classification of mergers in SMBHBs}

We now classify the different outcomes of the SMBHB-CO interactions with the aid of the stability map introduced in the previous section and visualized in Figure \ref{fig:stability}. In general, when $S_\mathrm{in}<0$, the tidal force exerted by the central SMBH is large enough to break apart the CO binary on a short timescale. {Those systems have been excluded from the initial conditions as we discussed above. Therefore, the legend 'decoupled' in \fig{fig:stability} indicates binary disruption from interaction with the secondary massive BH (i.e., disruption directly by the secondary massive BH or by the primary massive BH but as a result of a perturbation by the secondary massive BH)}. The right part of the stability map, where $S_\mathrm{in}>0$, is where all the interesting action occurs.  Here, the interaction can lead to different flavors of ejected and merging systems, as we now explore in detail. 

\subsubsection{Ejected mergers}

In the stability map where $S_\mathrm{out}<0$, the outer triple is unstable, indicating that the influence of $\mbtwo$ severely perturbs the orbit of the centre of mass of the CO binary. In fact, in this case the CO binary can be significantly affected by close encounters with $\mbtwo$, causing its ejection from the SMBHB sphere of influence with high velocity. Due to energy and angular momentum exchange with $\mbtwo$, ejected CO binaries can get tightened enough to merge within a Hubble time. { Depending on the ejection velocity and on the depth of the potential well of the host galaxy, these CO binaries can merge in the outskirts of the dark matter halo of their host galaxy, or even in the IGM.} 

This kind of merger event has two prerequisites: the CO binaries have to be ejected by the SMBHB (to produce HVBs), and the ejected CO binaries need to be tight enough to merge in a Hubble time due to GW radiation. To produce the HVBs, their ejection by the SMBHB must be efficient. This requires that the semi-major axis of the SMBHB is at least as small as $a_0$, and that the number density of CO binaries around  $\mbtwo$ is sufficiently large. At the same time, $S_\mathrm{out}$ needs to be smaller than zero to ensure that the CO binaries have a close encounter with $\mbtwo$, while $S_\mathrm{in}$ needs to be larger than zero to ensure that the CO binaries will not be disrupted by $\mbone$. To get compact HVBs that will merge within a Hubble time, the CO binaries need either to be sufficiently hard before the ejection or sufficient transfer of energy and angular momentum must occur from the binary's orbit to its centre of mass orbit around the SMBHB. Detailed descriptions of these processes can be found in \citet{2019MNRAS.482.3206W}. The distribution of ejected merger events in the stability map is shown in \fig{fig:stability} for all the simulated CO binary species.

\subsubsection{Quick mergers (i.e., collisions)}
In the inner triple system, LK cycles can directly create collisions between the components of the compact object binary if the eccentricity becomes sufficiently high. The typical (quadrupole) timescale for such an excitation cycle is \citep[e.g.,][]{1962P&SS....9..719L,1962AJ.....67..591K},
\be\label{eq:LK}
\begin{split}
\tau_\mathrm{LK,in} &\sim \frac{1}{n_\star}\bigg( \frac{\msone+\mstwo}{\mbone}\bigg) \bigg( \frac{\acm}{\astar}\bigg)^3(1-\ecm^2)^{3/2}\\
&=1.3\Myr\bigg( \frac{\mbone}{M_6}\bigg)^{-1}\bigg( \frac{\msone+\mstwo}{\solarm}\bigg)^{1/2}\\
&\times \bigg( \frac{\acm}{0.1\pc}\bigg)^{3}\bigg( \frac{\astar}{\au}\bigg)^{-3/2}(1-\ecm^2)^{3/2}\,,
\end{split}
\ee
 where $n_\star=\sqrt{G(\msone+\mstwo)/\astar^3}$ is the mean motion of the compact object binary. The strength of the next higher order LK effect (i.e., the octupole effect) is described by \citep[][]{2013MNRAS.431.2155N} 
\be\label{eq:oct_LK}
\epsilon_\mathrm{oct} = \frac{|\msone-\mstwo|}{\msone+\mstwo}\frac{\astar}{\acm}\frac{\ecm}{1-\ecm^2}\,.
\ee
Since the stability criterion in the inner triple requires the ratio ${\astar}/{\acm}$ to be extremely small, the CO binary will not be disrupted by the SMBHB and the octupole effect is usually negligible. 

To maintain the LK oscillations in the inner triple, the perturbations from the SMBH need to be stable. This requires that the centre of mass orbit of the CO binary will not change significantly during the inner LK cycles. Thus, the outer triple system also needs to be stable such that the perturbation from  $\mbtwo$ will not break the centre of mass orbit of the CO binary. In such a system, for which $S_\mathrm{in}>0$ and $S_\mathrm{out}>0$, we find a relatively short corresponding value for $\tau_{LK,in}$, If and hence we frequently detect collisions in the simulations. Merger events from this channel occur on a short time scale in the simulations, especially for CO binaries with component stars that have the largest radii (i.e., WD-WD binaries). Thus, throughout this paper, we call this kind of merger 'quick mergers'.

\begin{figure*}
    \includegraphics[width=\textwidth]{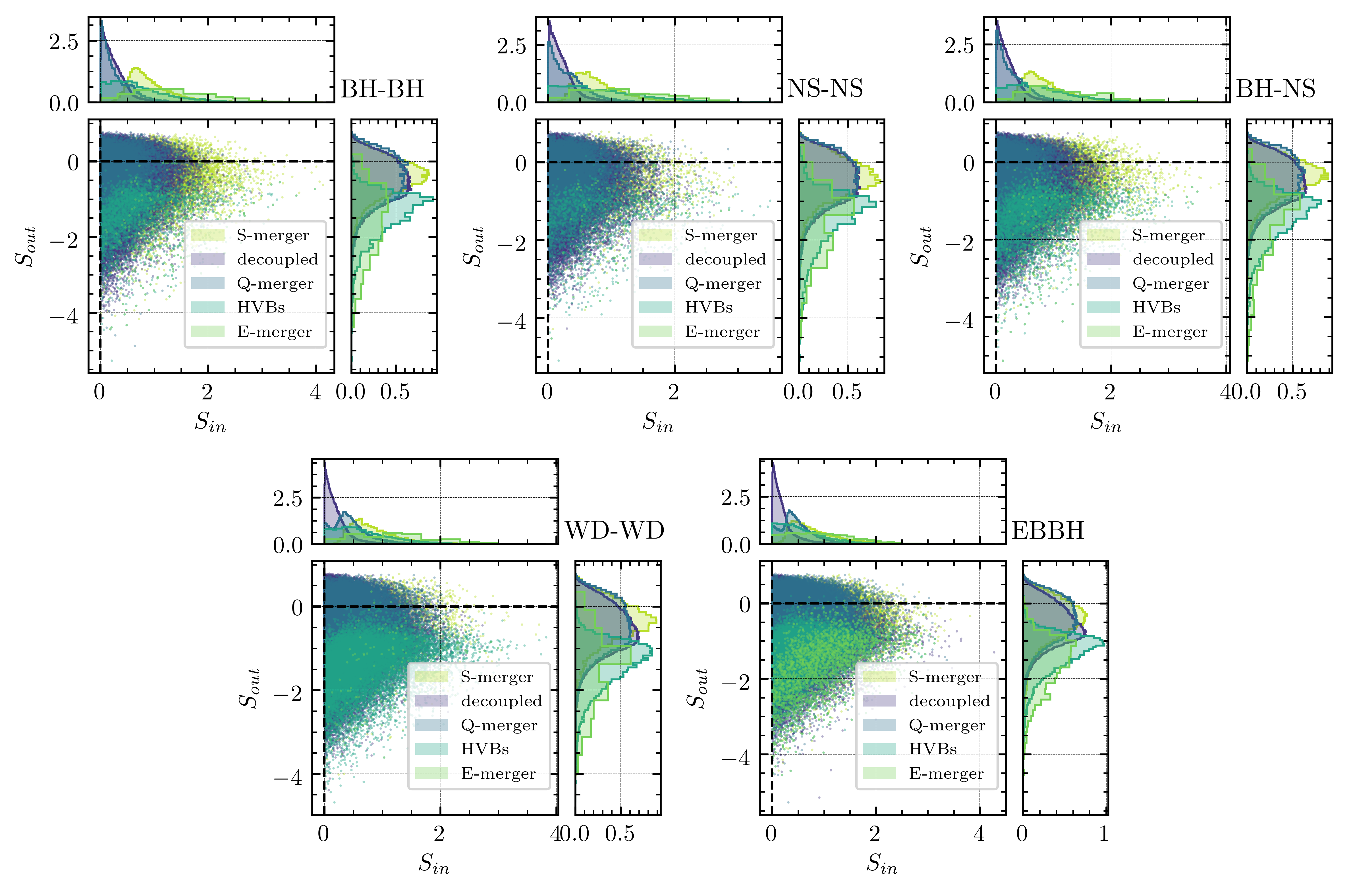}
     \caption{Distribution of the possible outcomes of our simulations in terms of the stability map. \textit{Top Left} panel: BH-BH, \textit{Top middle} panel: NS-NS, \textit{Top right} panel: BH-NS, \textit{Bottom left} panel: WD-WD. \textit{Bottom right} panel: EBBH. The marginalized distributions are normalized to 1 for visualization, thus they do not reflect the true event numbers of each event set. Also note that the ejected-merger is a subset of HVBs.}
    \label{fig:stability}
\end{figure*}

\fig{fig:stability} shows the quick mergers in the stability map ($S_\mathrm{in}$,\,$S_\mathrm{out}$). We can clearly see that for different CO binaries (i.e., WD-WD, BH-BH and NS-NS), almost all quick mergers occur in the upper right quadrant where $S_\mathrm{in}>0$ and $S_\mathrm{out}>0$. As expected, for the WD-WD binaries, the distribution for $S_\mathrm{in}$ is more dispersed. This is because the WD-WD binaries need only relatively low eccentricities to merge due to their larger radii, which requires only weaker LK oscillations that correspond to larger values for $S_{in}$ (larger ${\acm}/{\astar}$ in \eqn{eq:LK}). A signature of quick merger events is that they have non-negligible eccentricities in the LIGO band. This is because the CO binaries directly collide in our simulations before circularization can occur due to GW radiation. Note, however, that LK oscillations around a single SMBH can also lead to eccentric LIGO events \citep[e.g.][]{2018ApJ...860....5G}; therefore, eccentric mergers should not be considered signposts of SMBHBs.

\subsubsection{Slow mergers}

Some CO binaries neither experience ejections by $\mbtwo$ nor are subject to strong LK cycles. {This occurs for a subset of systems located within the stable region in our stability map} where $S_\mathrm{in}>0$ and $S_\mathrm{out}>0$.  Alternatively, they could remain bound to the SMBHB after a close encounter with $\mbtwo$ in the region where $S_\mathrm{in}>0$ and $S_\mathrm{out}<0$. CO binaries in the region where $S_\mathrm{in}>0$ and $S_\mathrm{out}>0$ which experience negligible LK effects are in the ideal environment to allow for binary stellar evolution (BSE) to occur. Thus, some tight CO binaries could form from BSE via common envelope evolution (for example), and then merge due to GW radiation within {the cosmological time elapsed since their formation, i.e. the lookback time $\tau_{\rm lb}$}. CO binaries in the region where $S_\mathrm{in}>0$ and $S_\mathrm{out}<0$ which remain bound to the SMBHB after a close encounter with the secondary $\mbtwo$ tend to become harder post-encounter. Therefore, a fraction of CO binaries in this region of phase space will be pushed into the merger band of LIGO by close encounters with $\mbtwo$. 

Despite the specific formation channel (BSE or encounter) for the tight CO binaries observed in the galactic centre, our simulations suggest a high formation rate if an IMBH is indeed present. Unlike quick mergers that are directly detected in the simulations, the 'slow' merger events will undergo significant circularization before merger due to GW radiation. Therefore, the eccentricity of the merger events for this channel will be completely negligible once in the LIGO band. Slow merger events are also shown in the stability map of \fig{fig:stability} (denoted with "S-merger").

\subsection{Outcomes of Scattering experiments}
\subsubsection{Relative fraction of HVBs}

\begin{figure*}
    \includegraphics[width=\textwidth]{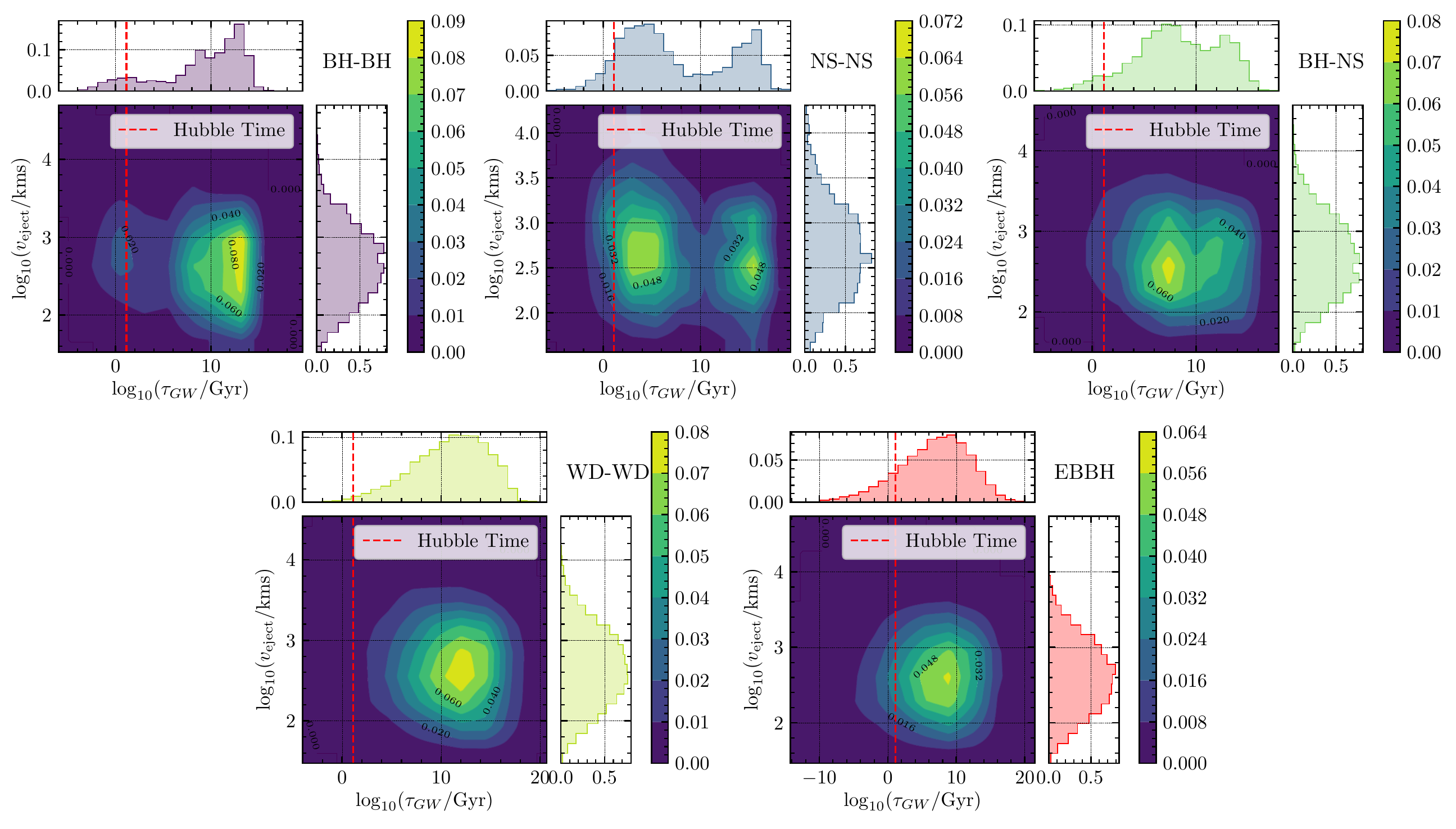}
     \caption{Contour plots of the 2D PDF $\int f_{\hvb}(m_\chirp,v_\ejct,\tau) \df m_\chirp$ of HVBs, i.e. the normalized differential distribution of the HVBs as a function of $v_\ejct$ and $\tau$. \textit{Top Left} panel: BH-BH, \textit{Top middle} panel: NS-NS, \textit{Top right} panel: BH-NS, \textit{Bottom left} panel: WD-WD. \textit{Bottom right} panel: EBBH.}
    \label{fig:TV-dist}
\end{figure*}

From our scattering experiments, {for each value of} $\mbone$ and $q$, the distribution of HVBs can be expressed as
\begin{equation*}
\frac{\df^4 N_\hvb}{\df\astar\df\estar\df m_\chirp\df v_\ejct}\,,
\end{equation*}
where $\astar$, $\estar$, $m_\chirp$ and $v_\ejct$ are the semi-major axis, eccentricity, chirp mass and ejection velocity of the HVB, respectively. All these quantities are directly obtained from the scattering experiments.

After the ejection, due to GW radiation, some HVBs will merge within an Hubble time. The merger time scale due to GW radiation is given by \citep[][]{1964PhRv..136.1224P}
\be
\tau_\gw(m_\chirp,\astar,\estar) = \frac{3}{85}\frac{c^5\astar^4(1-\estar^2)^{7/2}}{G^3\msone\mstwo(\msone+\mstwo)}\,.
\ee
Since not all HVBs will merge within {$\tau_{\rm lb}$}, we need to select only those HVBs whose merger time $\tau_\gw$ is shorter than  {$\tau_{\rm lb}$}. Previous works \citep[][]{2019MNRAS.482.3206W} have shown that from the close interaction with the SMBHB, many of the ejected binaries tend to become harder after the ejection due to energy and angular momentum exchange between the IMBH and the compact binary. Thus, we expect a significant fraction of the ejected HVBs to rapidly escape the SMBHBs and eventually merge (within $\tau_{\rm lb}$) in the dark matter halo of their host galaxy or even in the IGM. To compute the distribution of merger times due to GW emission, $\tau_\gw$, we have to convert the distribution of HVBs to the $(m_\chirp,v_\ejct,\tau)$-space by integrating over the parameter $\tau$,
\be
\begin{split}
&f_{\hvb}(m_\chirp,v_\ejct,\tau)=\frac{\df^3 N_\hvb}{\df m_\chirp\df v_\ejct \df \tau}\\
&=\frac{\df}{\df\tau}\oiint_{\tau_\gw(m_\chirp,\astar,\estar)<\tau}\frac{\df^4 N_\hvb}{\df\astar\df\estar\df m_\chirp\df v_\ejct}\df\astar\df\estar\,.\\
\end{split}
\ee

Figure \ref{fig:TV-dist} shows  $f_{\hvb}(m_\chirp,v_\ejct,\tau)$, 
which is the 2D-PDF of the hypervelocity binaries as a function of the ejection velocity $v_\ejct$ and the GW merger time $\tau_{\rm GW}$. Clearly, for each type of CO binary, we get some HVBs for which the final velocity is high enough relative to the local escape speed that these objects are expected to escape from the sphere of influence of their host SMBHBs.  After applying a specific galactic potential, one can compute whether or not a given binary will merge within the dark matter (DM) halo of their host galaxy or the IGM. { Detailed calculations regarding the cosmic rate of this type of merger will be discussed in \sect{sec:cosmic}.}

\subsubsection{Relative fraction of quick merger events}

\begin{figure*}
\includegraphics[width=\textwidth]{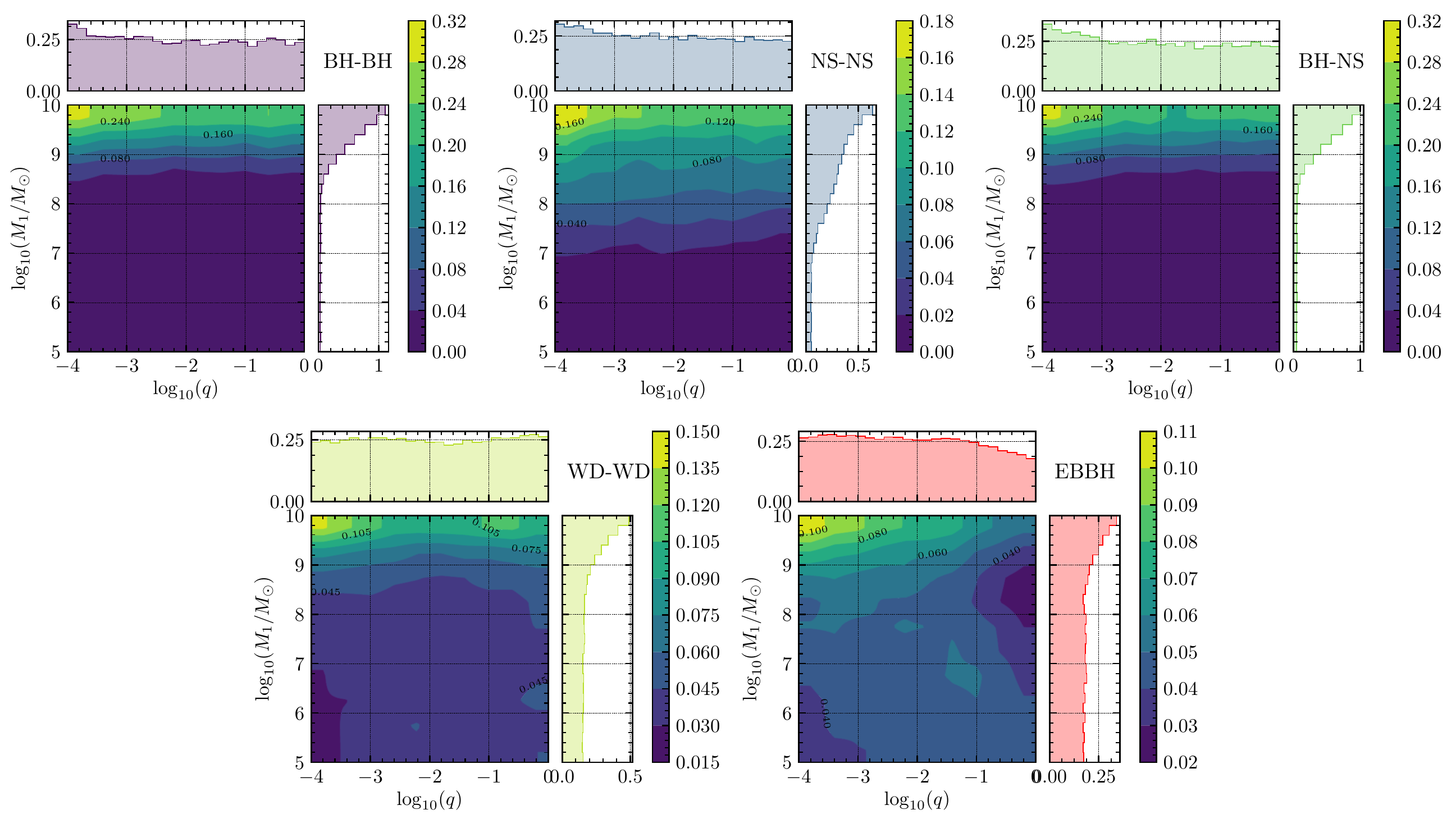}
\rule{\textwidth}{0.02cm}
\includegraphics[width=\textwidth]{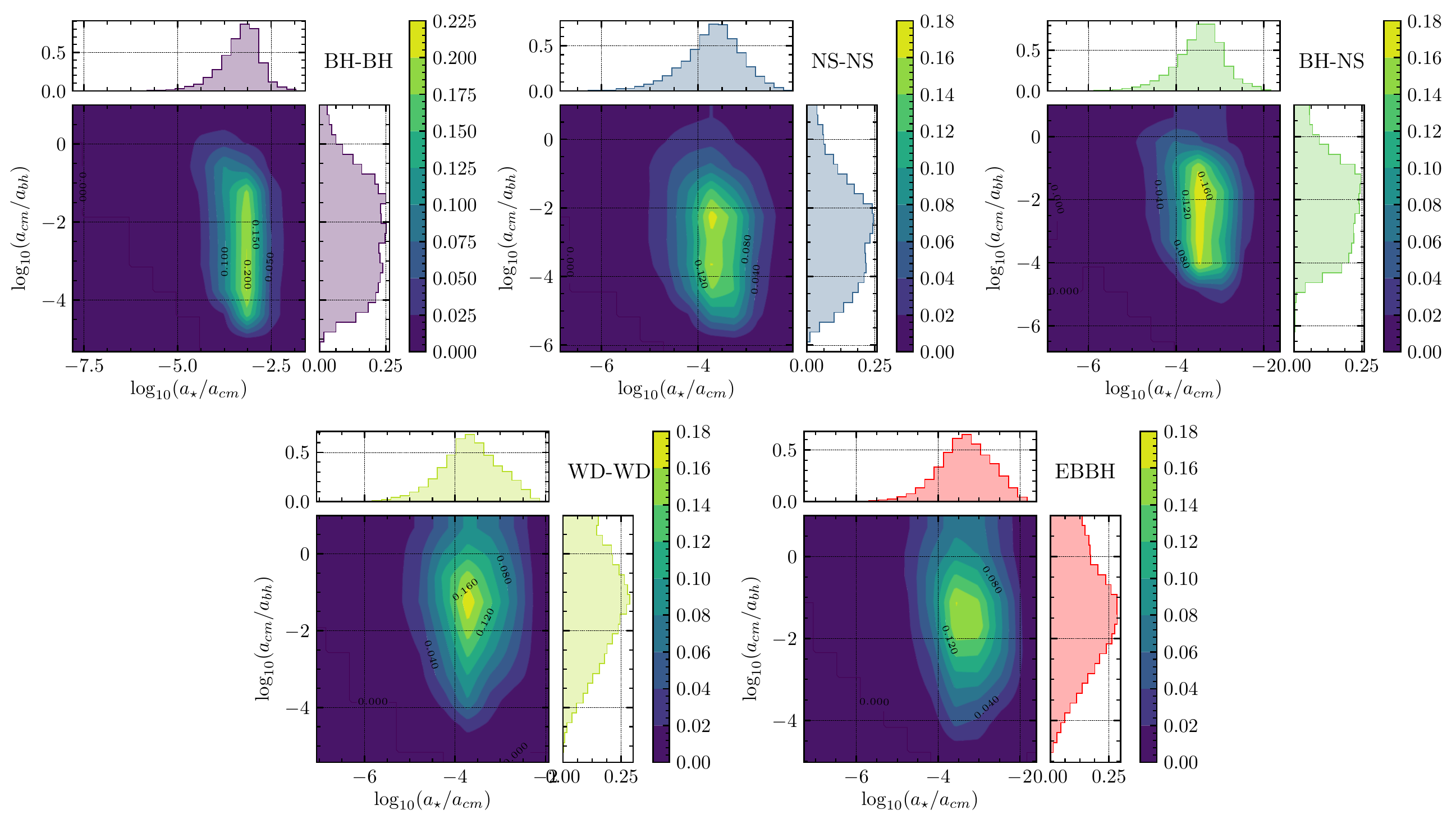}
     \caption{The 2D PDF of the relative fraction of quick mergers in the $M_1$-$q$ (top series of panels) and $\acm/\abh$-$\astar/\acm$ (bottom series of panels) planes. \textit{Top Left} panel: BH-BH, \textit{Top middle} panel: NS-NS, \textit{Top right} panel: BH-NS, \textit{Bottom left} panel: WD-WD. \textit{Bottom right} panel: EBBH.}
    \label{fig:quick_rel}
\end{figure*}

Similarly, we can examine the properties of quick merger events. Those are shown in \fig{fig:quick_rel}. The upper panels show, for the various types of CO binaries, the distribution 
\begin{equation*}
\frac{1}{N_\collision}\frac{\df^2 N_\collision}{\df\nm\df\nq}\,,
\end{equation*}
 {which indicates that the relative rate of quick mergers is independent of the mass ratio of the SMBHB and has a large probability for SMBHBs with large $\mbone$.}
  {Indeed, the quick merger rate depends on the time scale of the LK oscillations from the primary massive BH which is more sensitive to the location of the CO binaries in the SMBHB potential well ($\acm/\abh$), as well as the mass of $\mbone$.} The bottom panels show the distribution 
 \begin{equation*}
 \frac{1}{N_\collision}\frac{\df^2 N_\collision}{\df\ecm\df\log \acm}\,.
 \end{equation*}
When BH-BH, NS-NS and BH-NS binaries are involved, quick mergers tend to occur when $\acm/\abh\lesssim 10^{-1}$. To produce quick mergers, in fact, the LK-effect in the inner triple must be very efficient. The time scale for the leading quadrupole and octupole LK-effects are given by, respectively, \eqn{eq:LK} and \eqn{eq:oct_LK}. In this region, the strong LK oscillations efficiently drive the CO binary orbital eccentricity to extremely high values and cause a direct collision. The bottom two panels show that quick mergers in WD-WD binaries and EBBHs can also frequently occur when $a_{\rm cm}/a_0>1$. This is because WDs are much larger in size than either NSs or BHs, and binaries containing them can merge at lower eccentricities.  EBBHs themselves are much more eccentric, and thus need less help from LK oscillations. Therefore, only a moderate LK effect is required, which can occur farther from the central SMBH.

From our previous stability analysis, we know that quick mergers correspond mostly to the region $S_\mathrm{in}~>~0$ and  $S_\mathrm{out}~>~0$, for which both the inner and outer triple systems form stable hierarchical systems. In \fig{fig:stability}, we see that, as $S_\mathrm{in}$ approaches zero from above (indicating that the perturber comes closer to the compact object binary) the quick merger rate increases significantly. This is explained by the stronger inner LK effect due to a closer perturber. As $S_\mathrm{in}$ crosses zero, the perturbation from $\mbone$ becomes so strong that the inner triple becomes unstable, leading to binary disruption.  With that said, this is not seen in the stability map directly due to the 'hard cut' imposed in our initial condition at $S_\mathrm{in}=0$. Note that, since the WD radii are much larger than those of NSs and BHs and the EBBHs are very eccentric from the beginning, quick mergers of binary WD-WD and EBBHs require weaker LK oscillations, which is reflected by a wider dispersion in $S_\mathrm{in}$ in the bottom panels of Figure \ref{fig:stability}. 

Note that, in our simulations, the dissipative 2.5PN term is not included {due to the fact that the computation becomes prohibitively expensive for very tight CO binaries}. This narrows the eccentricity window in our collision detections. Eccentricities lower than what required for collisions to occur (but still extremely high) can generate quick mergers due to the high efficiency of GW radiation at very high $\estar$. These GW-induced mergers could have non-negligible eccentricities in the LIGO band, as was found in \citet[][]{2018PhRvD..97j3014S, 2012ApJ...757...27A}. Therefore, the quick merger rate obtained from our simulations corresponds to a lower limit.

\subsubsection{Relative fraction of slow merger events}

As for {ejected} mergers, we obtain the relative fraction of slow merger events as described in the following section.

\begin{figure*}
    \includegraphics[width=\textwidth]{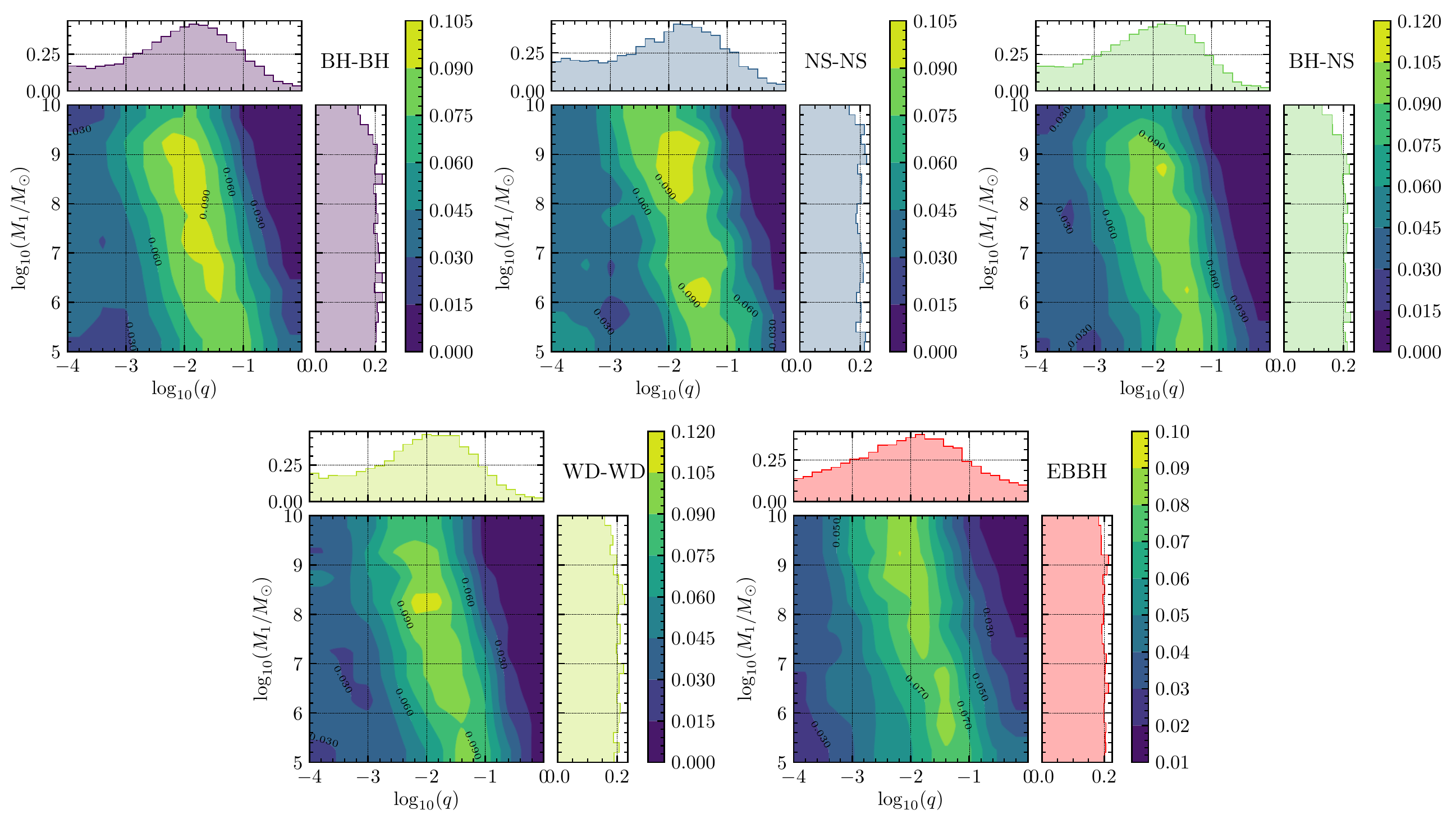}\\
    \rule{\textwidth}{0.02cm}
    \includegraphics[width=\textwidth]{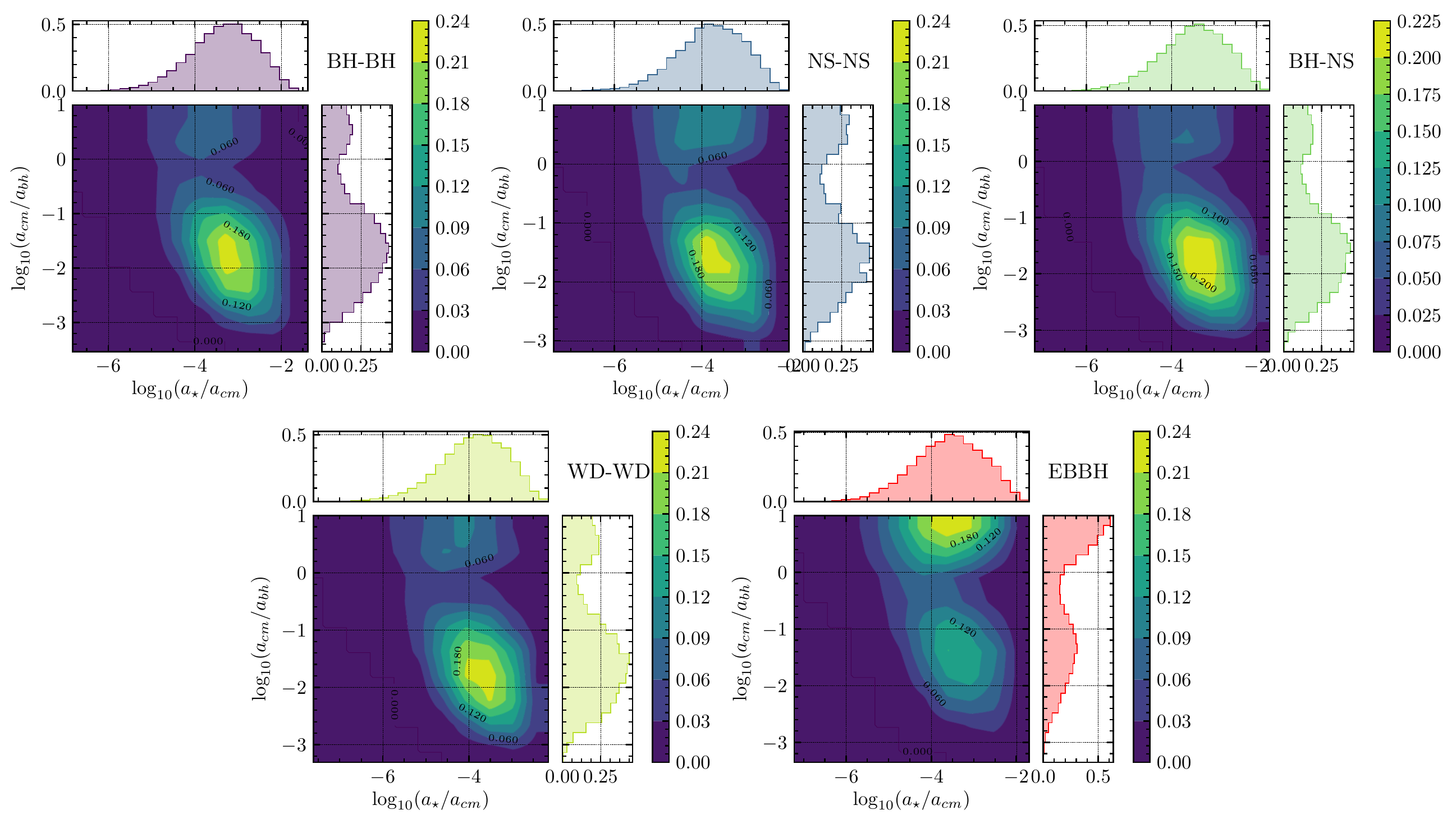}\\
     \caption{Same as \fig{fig:quick_rel} but for the relative fraction of 'slow mergers'. \textit{Top Left} panel: BH-BH, \textit{Top middle} panel: NS-NS, \textit{Top right} panel: BH-NS, \textit{Bottom left} panel: WD-WD. \textit{Bottom right} panel: EBBH.}
    \label{fig:incenter_rel}
\end{figure*}

From our scattering experiments, we obtain a number of unresolved interactions. After selecting those for which $\tau_\gw < \tau_\hubble$, we divide the results into two parts, namely $S_\mathrm{out}>0$ and $S_\mathrm{out}<0$. These two subsets correspond to the two different cases described above: 
binaries that could be formed from BSE and undergo long term LK oscillations from $\mbone$, or {tight binaries that remain bound to the SMBHB after close encounters with $\mbtwo$}. The two subsets can clearly be seen from the lower panels of \fig{fig:incenter_rel}, which shows the 2D-PDF of the slow merger rate in the ($\mbone$, $q$) and ($\ecm$, $\acm$) planes, namely
\begin{equation*}
\frac{1}{N_\imerger}\frac{\df^2 N_\imerger}{\df\nm\df\nq}\,
\end{equation*}
and
\begin{equation*}
 \frac{1}{N_\imerger}\frac{\df^2 N_\imerger}{\df\ecm\df\log \acm}\,.
\end{equation*}
The upper sets of panels in \fig{fig:incenter_rel} show that slow mergers from LK oscillations rarely occur for SMBHBs with mass ratios larger than 0.1.

The two subsets of events with $S_\mathrm{out}>0$ and $S_\mathrm{out}<0$ can clearly be recognized in the bi-modal distributions of the lower set of panels. The mode peaking at $\acm \sim [0.001, 0.1] \abh$ corresponds to stable systems with $S_\mathrm{out}>0$, whereas the mode extending at $\acm > \abh$ corresponds to unstable systems with $S_\mathrm{out}<0$. Note that since EBBHs merge much faster than the other four types of CO binaries in the region $\acm \sim [0.001, 0.1] \abh$, the distribution of slow mergers of EBBHs is dominated by the mode at $\acm > \abh$. 

\section{Cosmic rate of merger events from SMBHBs}\label{sec:cosmic}

{We now combine the outcomes of our scattering experiments with the SMBHB population extracted from the Millennium-II simulation, to estimate the cosmic merger rates of different types of mergers produced by SMBHB-CO binary interactions. Before proceeding, however, we shall stress an important caveat. Our simulations are meant to be representative of the dynamical interactions occurring as the SMBHB erodes the cusp of bound stars surrounding it as it first becomes bound. The cusp erosion process shrinks the binaries by a factor of $\approx 10$ starting from $a_0$, which is given by \eqn{eq:a0} if we assume that the cusp distribution follows an isothermal sphere. This might not be sufficient to efficiently drive the SMBHB to coalescence, and further shrinking, via the interaction of unbound stars scattered in the SMBHB loss cone, might be required. In general, shrinking by only a few e-folds is needed \citep[e.g.,][]{2010ApJ...719..851S} and each e-fold requires the interaction with a mass in stars of the order of $M_2$ \citep[see, e.g.,][]{2005LRR.....8....8M}. Since the erosion of the bound cusp ejects a mass in stars of order $\approx M_2$, the rates estimated in the following should be considered as conservative, and might be higher by up to a factor of a few.}

{We caution, however, that interaction with unbound binaries is inherently different than for bound binaries. Most stars, especially for SMBHB mass ratios close to unity, are either ejected in a single fly-by, or captured in relatively short-lived metastable orbits. LK oscillations are therefore irrelevant in this case, and the rate of quick and slow mergers should be strongly suppressed. On the other hand, CO binaries can still be ejected as HVBs, but the ejection velocities tend to be less extreme than those achieved in bound scattering. Unbound SMBHB-CO binary scattering experiments will eventually be needed to quantify the contribution of this late phase of the SMBHB evolution to the cosmic rate of ejected CO binary mergers. This is deferred to future work.}   
 
\subsection{Ejected merger rate}

From the scattering experiments, we have obtained the HVB distribution ${\df^3 N_\hvb}/({\df m_\chirp\df v_\ejct \df \tau})$. Note that this distribution varies with $\mbone$ and $q$, thus it is a function of $m_\chirp$, $\tau$ and $v_{\rm eject}$ but also of $\nm$ and $\nq$. After the ejection, a given HVB will escape from the galactic potential with initial velocity $v_\ejct$ and merge in the dark matter halo or in the IGM at a distance $R$ away from the host galactic centre.  This distance depends on the flight time $\tau$, and can be computed explicitly by assuming a specific galactic potential. In this work, we adopt the characteristic galactic potential of elliptical galaxies \citep[e.g.,][]{2008ApJ...680..312K,2014ApJ...793..122K},

\be
\begin{split}
\phi(r)&=\phi_\bh(r)+\phi_b(r) + \phi_h(r)\\
       &= -\frac{G(\mbone+\mbtwo)}{r} -\frac{GM_b}{r+r_b} - \frac{GM_h}{r}\ln\left(1+\frac{r}{r_h}\right)\,,
\end{split}
\label{eq:pot}
\ee
where $M_b$ is the mass of the bulge, $M_h$ is the mass of the dark matter halo, $r_b$ is the scale radius of the bulge and $r_h$ is the scale radius of the dark matter halo. The choice of the spherical potential in equation \eqref{eq:pot} is dictated by computational convenience, as it allows us to perform 1D calculations to compute the ejected CO binary trajectory. The addition of a disk component would produce a deflection of the ejected trajectories, without significantly affecting the overall distribution of distances travelled. We scale the variables $M_b$, $M_h$, $r_b$ and $r_h$ with the total mass of the SMBHB $\mbone(1+q)$ by means of the relations \cite[e.g.,][]{2013ARA&A..51..511K, 2015MNRAS.452.4013K}
\be
\begin{aligned}
&\frac{\mbone(q+1)}{10^9\solarm}&=&\,0.49\left(\frac{M_b}{10^{11}\solarm}\right)^{1.16}\,,\\
&\frac{M_h}{10^{13}\solarm}&=&\,0.30\left(\frac{M_b}{10^{11}\solarm}\right)\,,\\
&\frac{M_b}{10^6 M_\odot}\bigg(\frac{\mathrm{kpc}}{r_b}\bigg)^3&=&\, 4000\,,\\
&\frac{M_h}{10^6 M_\odot}\bigg(\frac{ \mathrm{kpc}}{r_h}\bigg)^3&=&\,125\,.
\end{aligned}
\ee

The ejection properties are recorded at a distance of $r\,\sim\,50\abh$, given the condition that the total energy of the compact binary must be positive. Thus, we can calculate the traveled distance $R=D(\mbone(1+q),\tau,v_\ejct)$ 
\be
\int_0^D\frac{dr}{\sqrt{2E_0-2\phi(r)- 2l_0^2/r^2}} =\int_0^{\tau_\gw} dt\,,
\ee
where $E_0$ and $l_0$ are, respectively, the total energy and total angular momentum per unit mass at the time of ejection. The distribution of HVBs as a function of mass, traveling time $\tau$ and distance from the galaxy centre at a merger distance $R$ can then be calculated as
\be
\begin{split}
&\frac{\df^3 N_\hvb}{\df m_\chirp\df \tau\df R}(\nq,\nm,m_\chirp,\tau,R)\\
&=\frac{\df}{\df R}\oint_{D(\mbone,q,\tau,v_\ejct) < R}\frac{\df^3 N_\hvb}{\df m_\chirp\df v_\ejct\df \tau}\df v_\ejct\\
\end{split}\,.
\ee
\begin{figure*}
    \includegraphics[width=\textwidth]{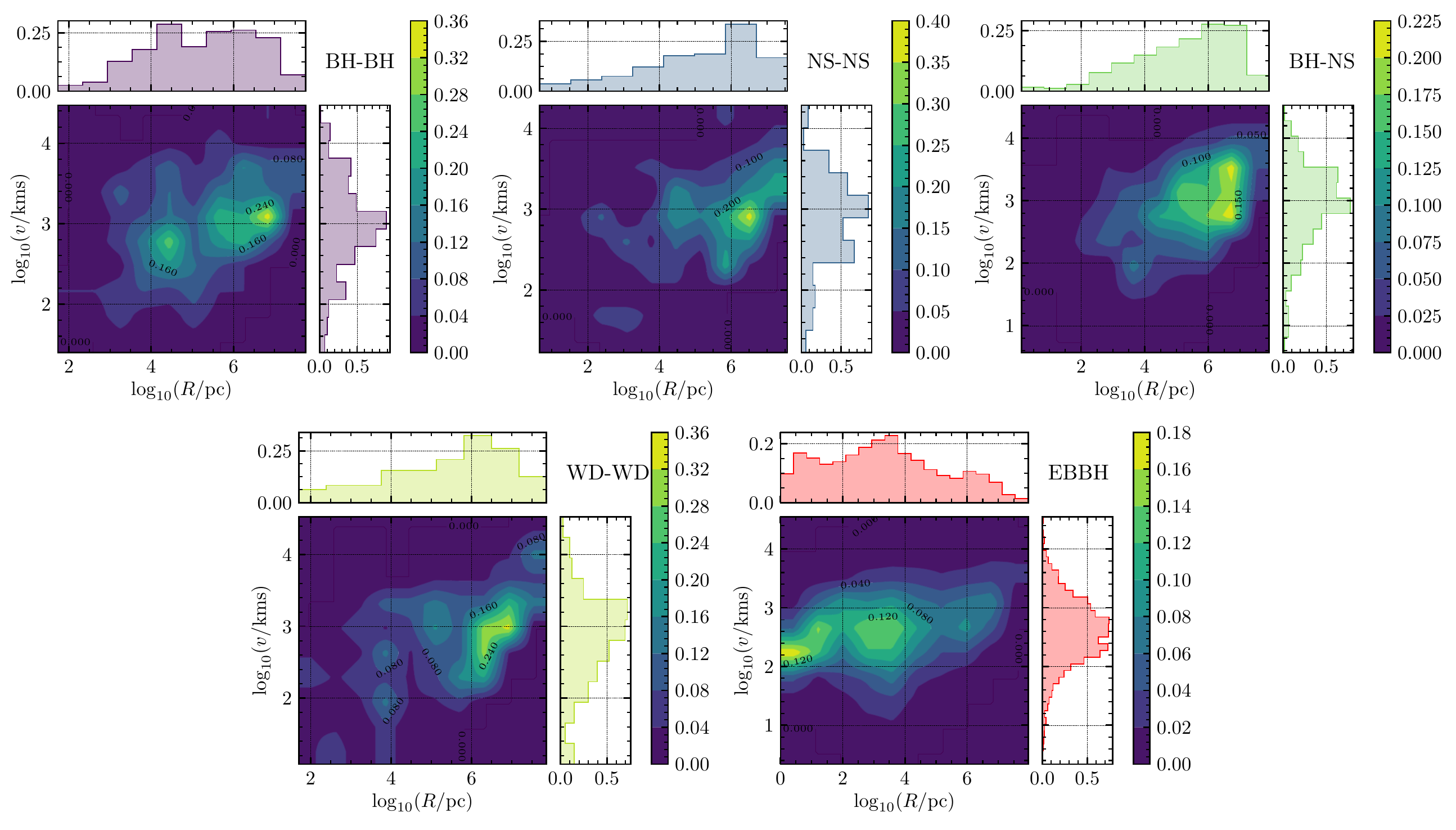}\\
     \caption{{The 2D-PDF of the ejected merger events as a function of the merger distance $R$ from their host galaxy,  as well as the corresponding CoM velocity when they  merge}.\textit{Top Left} panel: BH-BH, \textit{Top middle} panel: NS-NS, \textit{Top right} panel: BH-NS, \textit{Bottom left} panel: WD-WD. \textit{Bottom right} panel: EBBH. }
    \label{fig:RV}
\end{figure*}
\fig{fig:RV} shows the 2D-PDF of the distance $R$ and the corresponding velocity at that distance of the simulated HVBs. Note that not all ejected mergers can be observed within the age of the universe.

The results of our scattering experiments yield the time at which the SMBHB ejects the CO binaries, which corresponds to the redshift $z$. Merger events occur a time $\tau$ after the ejection. Therefore, there is a significant displacement of the merger redshift with respect to $z$. Only those events with $\tau\lesssim \tau_{\rm lb}$ can be observed; let us define the merger redshift to be $z^\prime=g(z,\tau)$, where
\be
\begin{split}
\tau &= \int_{z^\prime}^z \frac{\df t}{\df \bar{z}}\df \bar{z}\\
&= \frac{1}{H_0}\int_{z^\prime}^z\frac{\df \bar{z}}{(1+\bar{z})\sqrt{\Omega_M(1+\bar{z})^3+\Omega_k(1+\bar{z})^2+\Omega_\Lambda}}\,.\\
\end{split}
\ee
We solve this equation numerically to obtain the value of $z^\prime$ for each given $z$ and $\tau$. With this relationship, we can calculate the redshift distribution of the ejected mergers. Only mergers with corresponding redshift $z^\prime > 0$ can be observed today.

To compute the cosmic ejected merger rate in the rest frame of the Earth, we need to convolve the fraction of {HVBs} obtained from the scattering experiments with the cosmic distribution of compact binaries actively interacting with SMBHBs per unit time. As discussed in Section \ref{sec:intrate}, we compute the distribution of the actively interacting mass per unit time via  \eqn{eq:Mactive};
thus it is easy to obtain the cosmic distribution for the actively interacting binary number per unit time,
\begin{eqnarray}
\dot{n}_\act(z,\nq,\nm)&=&\frac{\df^4 N_\act}{\df z\df\nq\df\nm\df t}\\
&=&\frac{\df^4 (n_\bi M_\act)}{\df z\df\nq\df\nm\df t}\\
&=&n_{\bi}\dot{\mathcal{M}}_\act(z,\nq,\nm)\,,
\end{eqnarray}
where $n_\bi$ is the number of compact binaries (BHB, WDB, NSB) per unit mass in the galactic centre, which can be estimated via consideration of stellar evolution with a given IMF, and the binary fraction typical of galactic centres.

From our scattering experiments, {for each} value of $\mbone$ and $q$, we obtain the HVB distribution ${\df^3 N_\hvb}/({\df m_\chirp\df \tau\df R})$ as a function of the parameters $m_\chirp$, $\tau$ and $R$. Note that this distribution varies with $\mbone$ and $q$, thus it is not only a function of $m_\chirp$, $\tau$ and $R$ but also a function of $\nm$ and $\nq$. Therefore, the cosmic rate of HVBs can be calculated as
\be
\begin{split}
&\frac{\df^4N_\omerger}{\df z^\prime\,\df m_\chirp\,\df R\,\df t}(z^\prime, m_\chirp, R)\\
&=\theta(z^\prime)\frac{\df}{\df z^\prime}\int_{g(z,\tau)<z^\prime}\df\tau\df z\int\frac{\df^3 N_\hvb}{\df m_\chirp\df \tau\df R}\\
&\times\frac{\df^4 N_\act}{\df z\df\nq\df\nm\df t}\frac{1}{N_{\mathrm{exp}}(\nq,\nm)}\df \nm \df\nq\,,\\
\end{split}
\ee
where $N_{\mathrm{exp}}$ is the total number of scattering experiments performed for each $\mbone$ {and $\theta(z^\prime)$ selects out the merger events that can be observed to date ($z^\prime\ge 0$).}

\begin{figure*}
    \includegraphics[width=\textwidth]{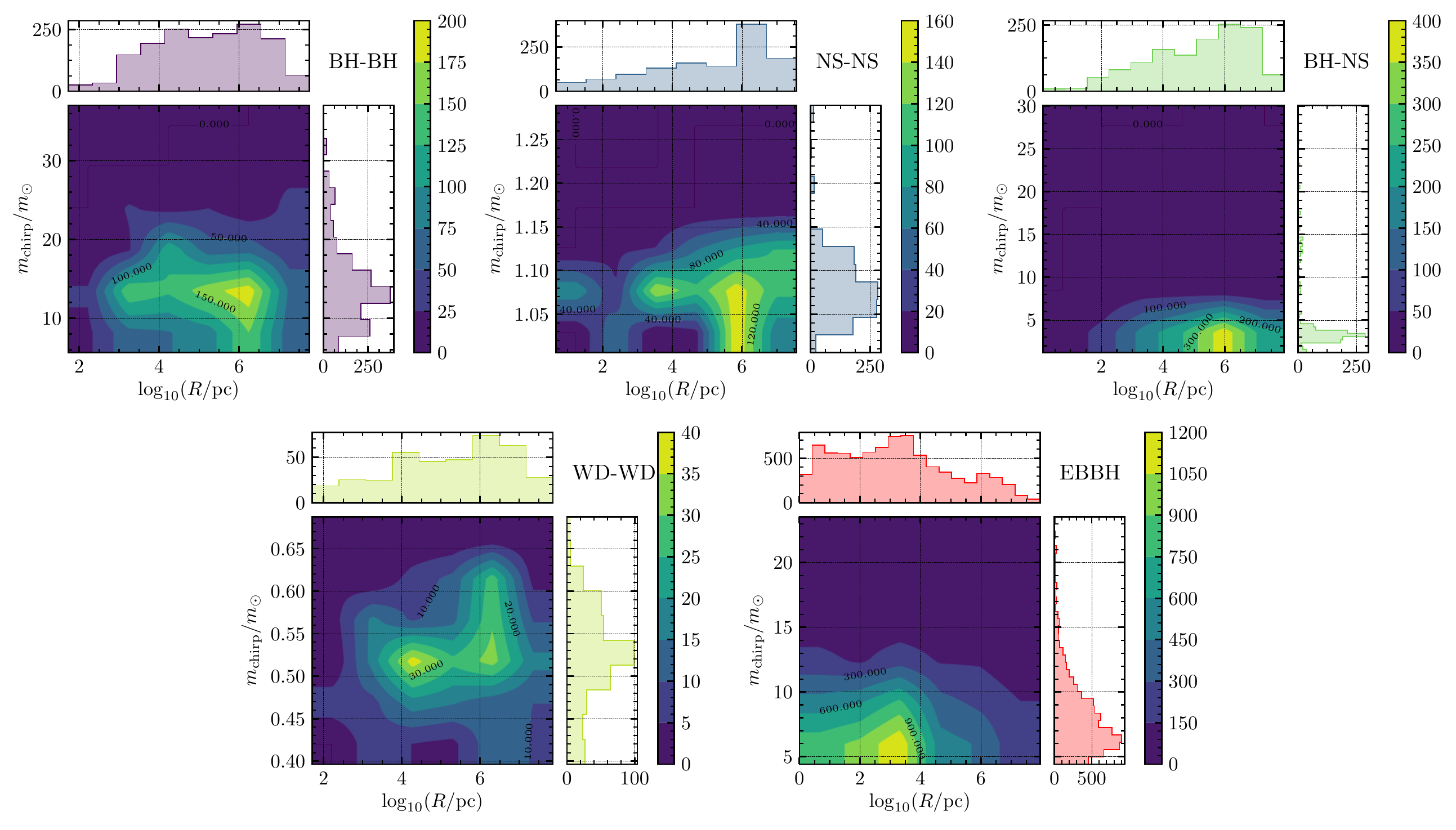}\\
    \rule{\textwidth}{0.02cm}
    \includegraphics[width=\textwidth]{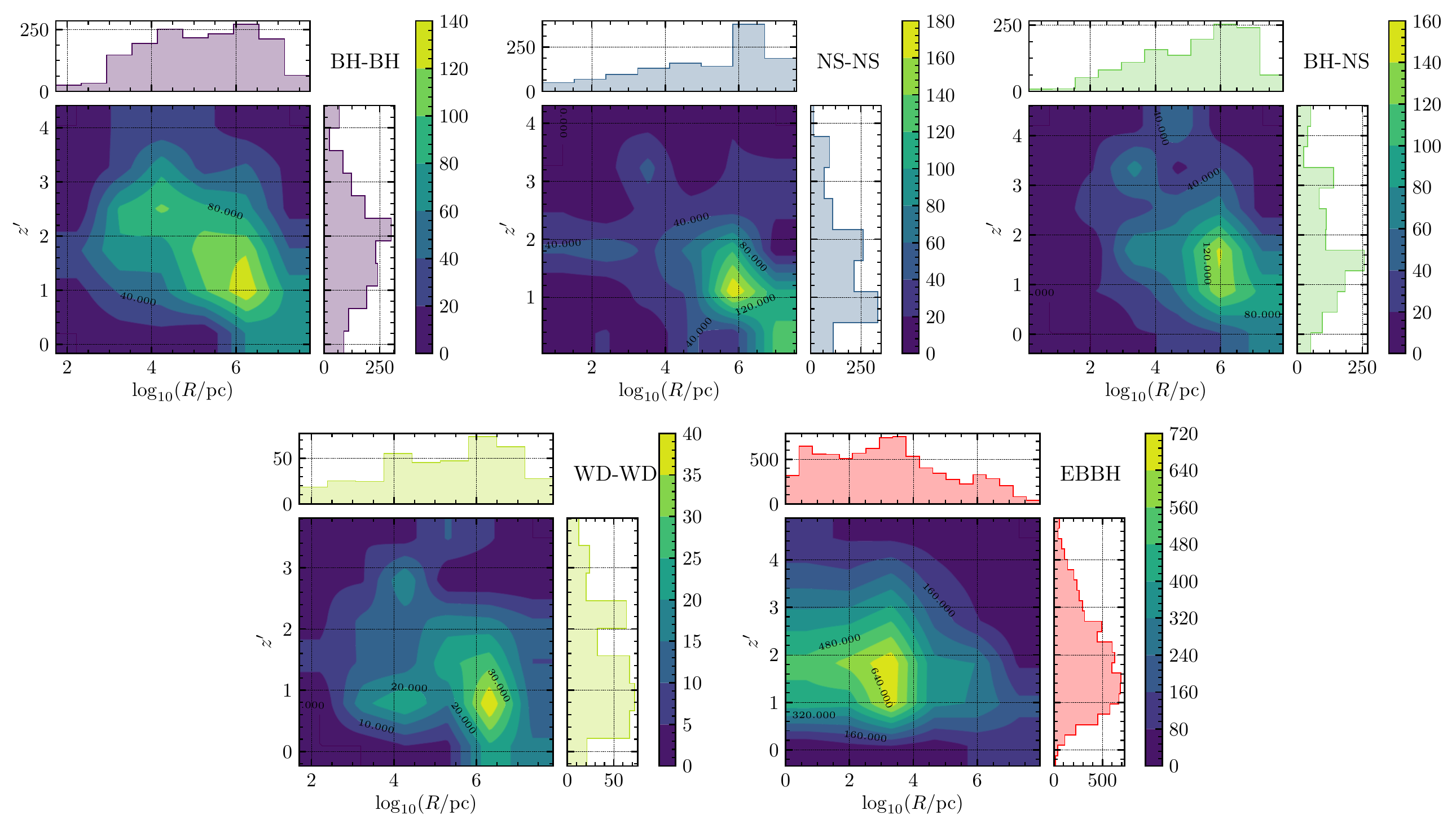}
     \caption{Cosmic distribution of ejected merger events and their dependences on the parameters $R$, $m_\chirp$ and $z^\prime$ normalized by the undetermined parameter $n_\mathrm{bi}$, where $n_\mathrm{bi}$ is the CO binary number per unit solar mass in the galactic centre. \textit{Top Left} panel: BH-BH, \textit{Top middle} panel: NS-NS, \textit{Top right} panel: BH-NS, \textit{Bottom left} panel: WD-WD. \textit{Bottom right} panel: EBBH.}
    \label{fig:cosmic_rate}
 \end{figure*}

 The upper panels of \fig{fig:cosmic_rate} show the distribution of the rate of cosmic '{ejected}' mergers {that can be observed up to date, i.e.  ${\df^3N_\omerger}/({\df m_\chirp\,\df R\,\df t})$}. For all types of CO binaries, we find merger events occurring both in dark matter halos and in the interstellar and intergalactic media ($\unit{kpc-Mpc}$). There is a peak in $R$ around $\unit{Mpc}$ for WD-WD, BH-BH, NS-NS and BH-NS binaries, indicating that those systems can merge deep in the IGM. Conversely, EBBHs preferentially merge at $R<10^{-2}\unit{Mpc}$, due to their much shorter GW radiation merger times. The bottom subplots of \fig{fig:cosmic_rate} show the distribution of the cosmic {ejected} merger rate ${\df^3N_\omerger}/({\df z^\prime\,\df R\,\df t})$. We see that most of the {ejected} mergers should be detected in the range $z^\prime \sim [0.5,3]$.

 By integrating ${\df^4N_\omerger}/({\df z^\prime\,\df m_\chirp\,\df R\,\df t})$ we get
 \be
 \Gamma_\omerger = \frac{\df N_\omerger}{\df t} = \int \frac{\df^4N_\omerger}{\df z^\prime\,\df m_\chirp\,\df R\,\df t}\df z^\prime \df R\df m_\chirp
 \ee
which is the {ejected} merger event rate up to redshift $5$. The numerical values of the merger rate can be found in \tab{tab:R}, where $n_\mathrm{wd-wd}$, $n_\mathrm{ns-ns}$, $n_\mathrm{bh-bh}$, $n_\mathrm{bh-ns}$ and $n_\mathrm{ebbh}$ are the binary number per unit solar mass in the galactic centre for WD-WD, NS-NS, BH-BH, BH-NS and EBBH, respectively. {In the table, a subset of WD-WD mergers for which the total mass $>1.4 M_\odot$ has been listed separately. We simply label these as Type Ia SNe.} 

To assess the detectability of emitted GWs, we must consider the signal-to-noise ratio (SNR) of the GW detectors that can detect the events out to $z^\prime=f(m_\chirp)$. With this, the detectable {ejected} merger event rate as a function of $R$ and $z^\prime$ can be derived from

\be
\frac{\df^3N_\omerger}{\df R\df z^\prime\df t}(R,z^\prime)=\int_{f(m_\chirp)<z^\prime}\frac{\df^4N_\omerger}{\df z^\prime\,\df m_\chirp\,\df R\,\df t}\df m_\chirp\,.
\ee

\subsection{Quick merger rate}

\begin{figure*}
    \includegraphics[width=\textwidth]{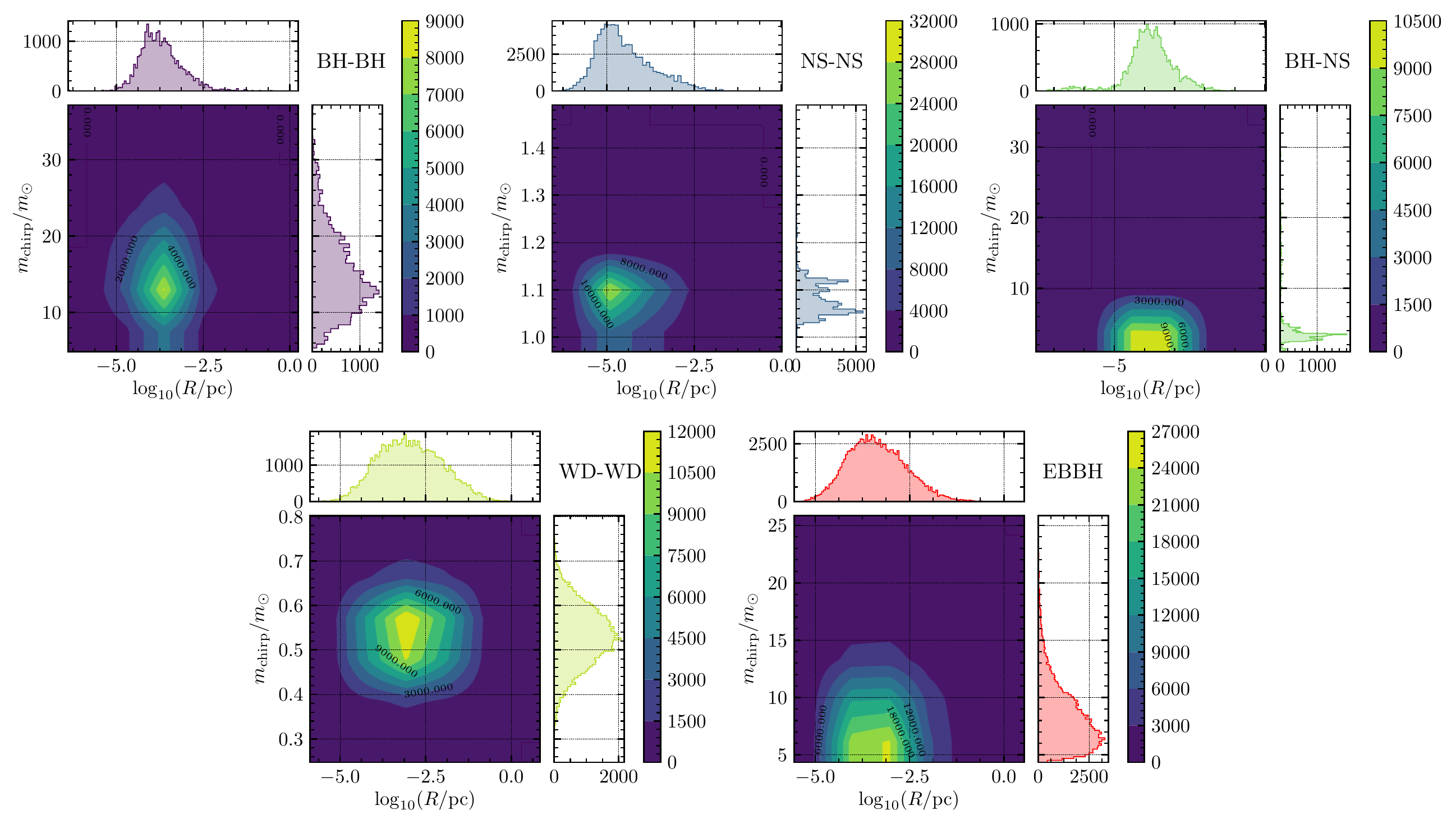}\\
    \rule{\textwidth}{0.02cm}
    \includegraphics[width=\textwidth]{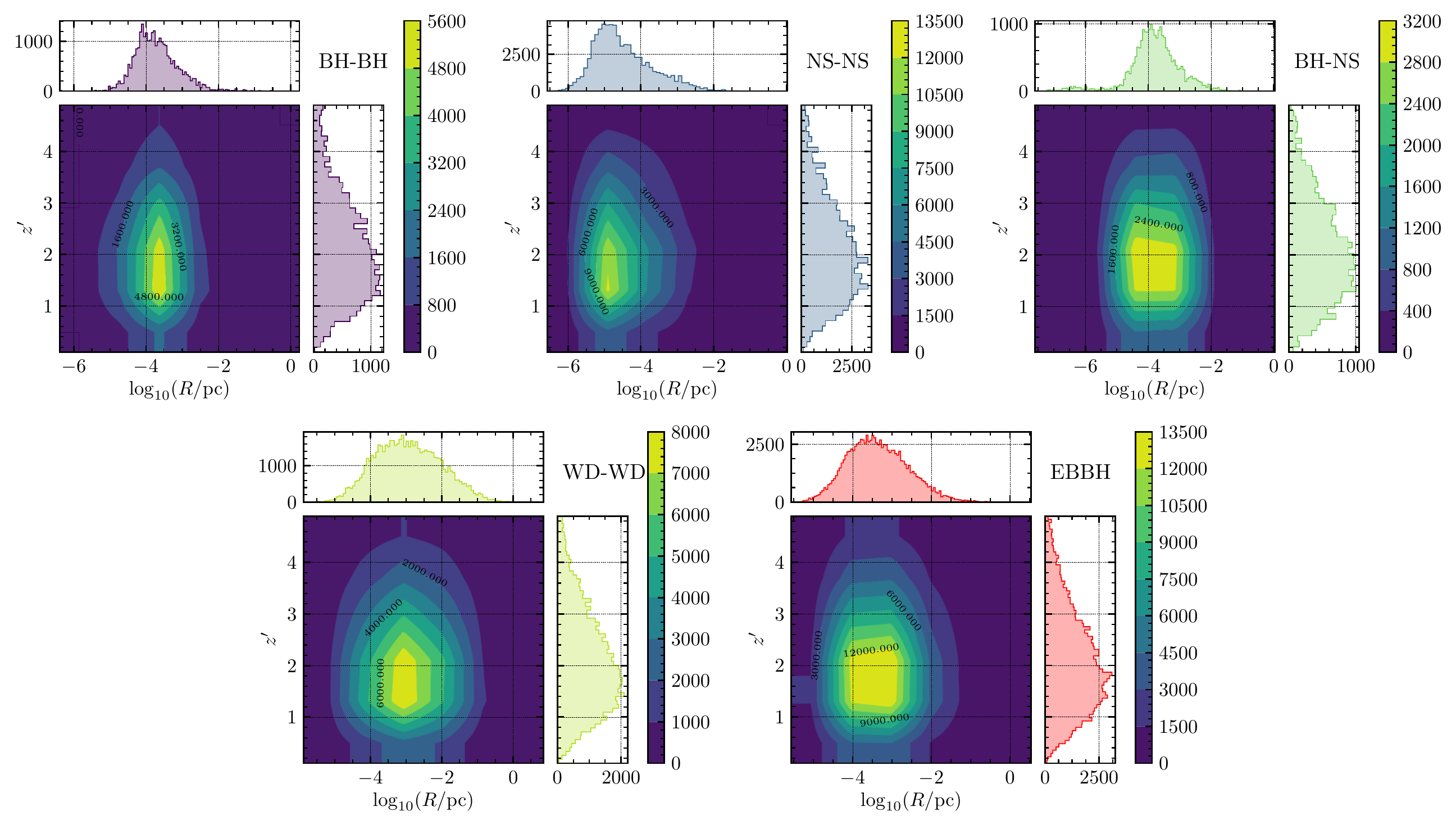}
     \caption{Cosmic distribution of quick merger events and their dependences on the parameters $R$, $m_\chirp$ and $z^\prime$ normalized by the undetermined parameter $n_\mathrm{bi}$, where $n_\mathrm{bi}$ is the CO binary number per unit solar mass in the galactic centre. \textit{Top Left} panel: BH-BH, \textit{Top middle} panel: NS-NS, \textit{Top right} panel: BH-NS, \textit{Bottom left} panel: WD-WD. \textit{Bottom right} panel: EBBH.}
    \label{fig:cosmic_quick_rate}
 \end{figure*}

Following the same procedure as for {ejected} merger events, we can calculate the cosmic rate of quick merger events as follows,
\be
\begin{split}
&\frac{\df^3N_\collision}{\df z^\prime,\df m_\chirp\,\df t}(z^\prime, m_\chirp)\\
&=\int N_{\collision}\frac{\df^4 N_\act}{\df z\df\nq\df\nm\df t}\frac{\delta(z-z^\prime)}{N_{\mathrm{exp}}(\nq,\nm)}\df \nm \df\nq\,,
\end{split}
\ee
where $N_{\collision}$ is verified to be independent of $\mbone$ and $q$ in \sect{sec:relfraction}.

\fig{fig:cosmic_quick_rate} shows the distribution of ${\df^3N_\collision}/({\df z^\prime,\df m_\chirp\,\df t})$;  the corresponding rates can be obtained by
 \be
 \frac{\df N_\collision}{\df t} = \int \frac{\df^3N_\collision}{\df z^\prime\,\df m_\chirp\,\df t}\df z^\prime \df m_\chirp
 \ee
 The numerical values of the merger rate can be found in \tab{tab:R}.

\subsection{Slow merger rate}

As for the quick mergers, we estimate the cosmic rate of {slow} mergers as
\be
\begin{split}
&\frac{\df^3N_\imerger}{\df z^\prime,\df m_\chirp\,\df t}(z^\prime, m_\chirp)\\
&=\int \frac{\delta(z-z^\prime)\df^4 N_\act}{\df z\df\nq\df\nm\df t}\frac{N_{\imerger}(\nm,\nq)}{N_{\mathrm{exp}}(\nq,\nm)}\df \nm \df\nq\,.
\end{split}
\ee

\begin{figure*}
    \includegraphics[width=\textwidth]{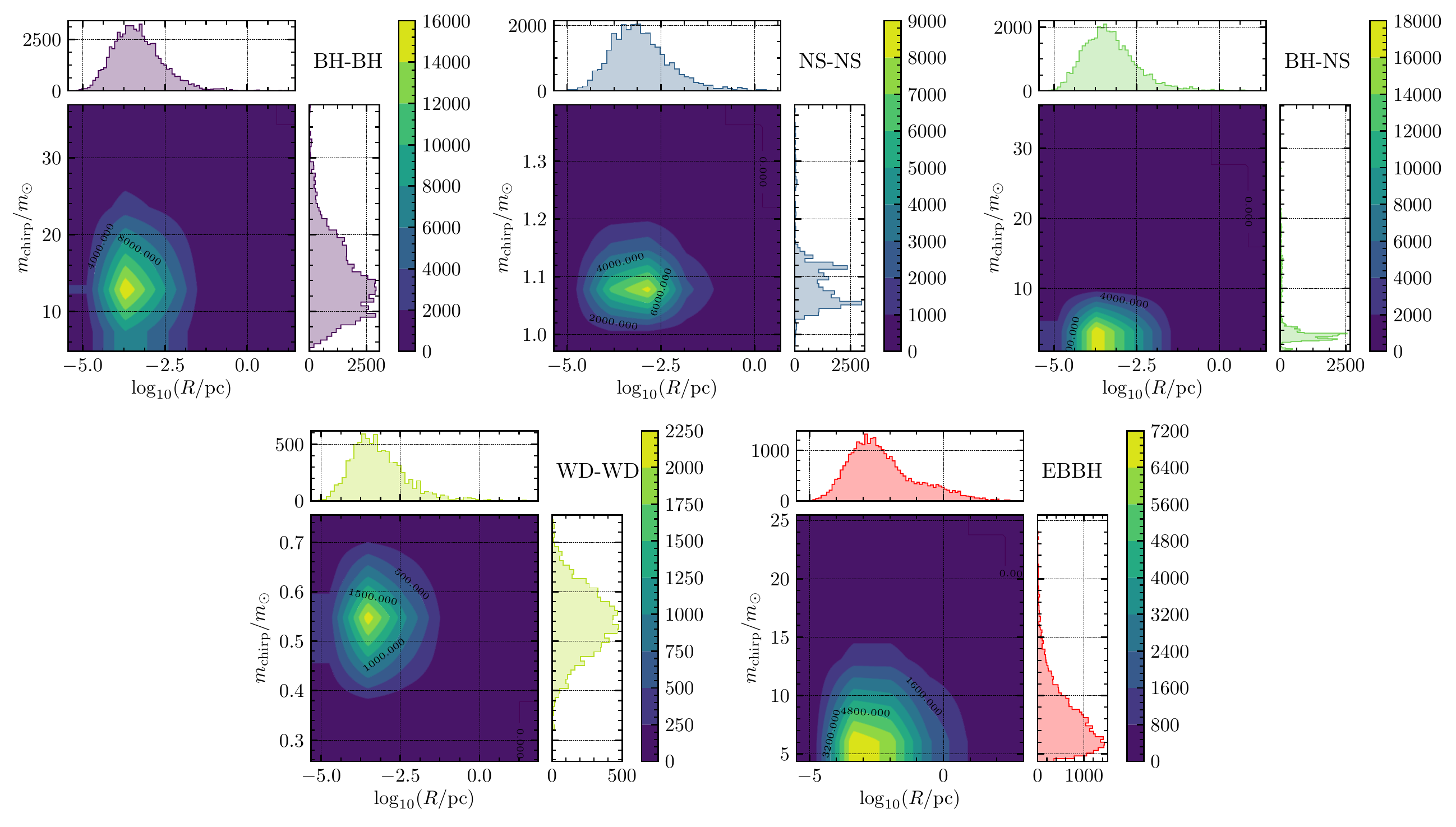}\\
    \rule{\textwidth}{0.02cm}
    \includegraphics[width=\textwidth]{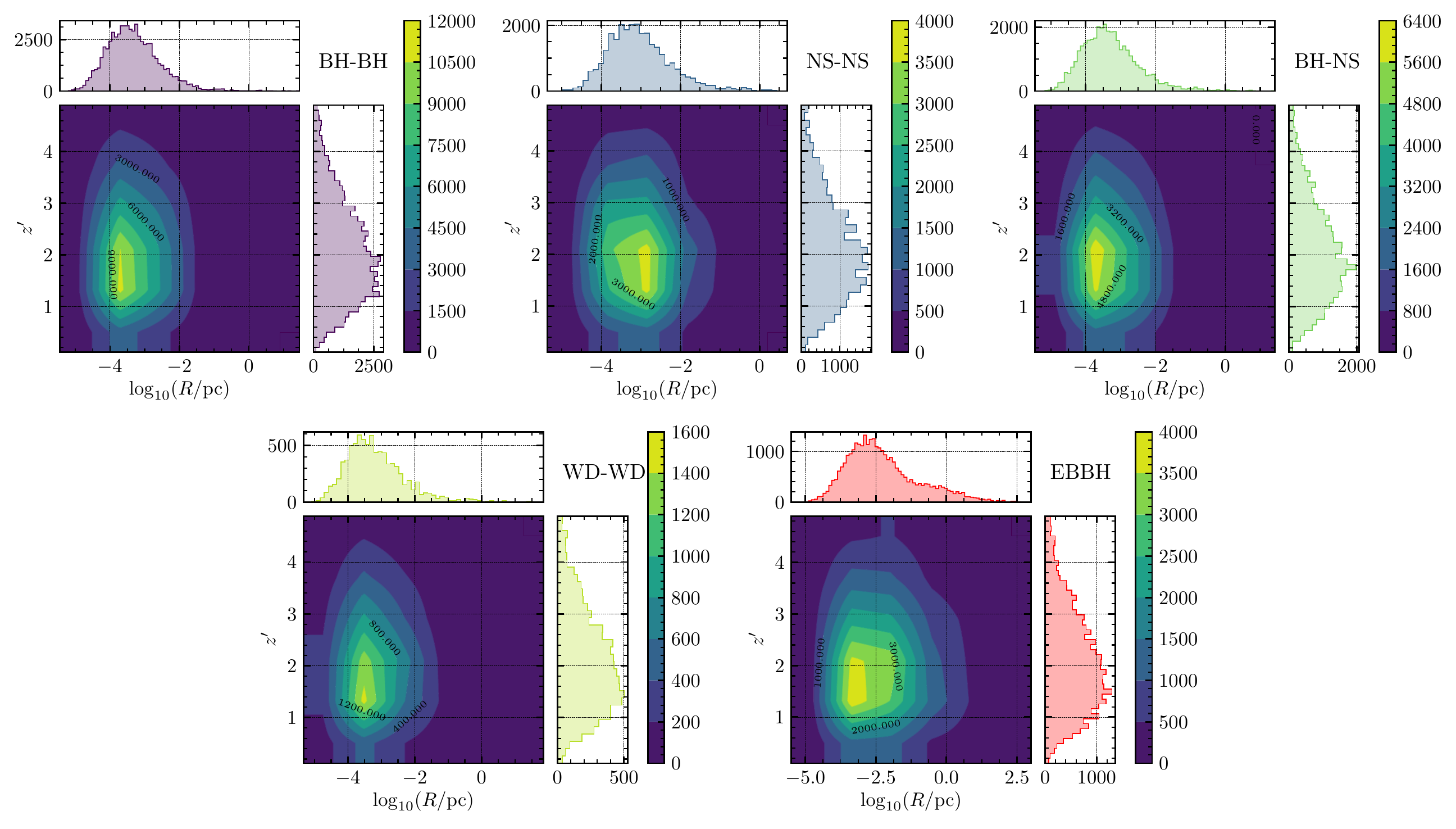}
     \caption{Cosmic distribution of slow merger events and their dependences on the parameters $R$, $m_\chirp$ and $z^\prime$ normalized by the undetermined parameter $n_\mathrm{bi}$, where $n_\mathrm{bi}$ is the CO binary number per unit solar mass in the galactic centre. \textit{Top Left} panel: BH-BH, \textit{Top middle} panel: NS-NS, \textit{Top right} panel: BH-NS, \textit{Bottom left} panel: WD-WD. \textit{Bottom right} panel: EBBH.}
    \label{fig:cosmic_in_rate}
 \end{figure*}

 \fig{fig:cosmic_in_rate} shows the distribution ${\df^3N_\imerger}/({\df z^\prime\,\df m_\chirp\,\df t})$, which is almost identical to the quick merger distribution. However, the non-negligible residual eccentricities for quick mergers determine the region in phase-space where the merger events tend to occur, as indicated by \fig{fig:incenter_rel} and \fig{fig:quick_rel}. The corresponding rates are computed via the relation
 \be
 \frac{\df N_\imerger}{\df t} = \int \frac{\df^3N_\imerger}{\df z^\prime\,\df m_\chirp\,\df t}\df z^\prime \df m_\chirp
 \ee
 The numerical values of the merger rate can be found in \tab{tab:R}.

\subsection{Merger distance from the galactic centre}
Due to differences in the surrounding environments, merger events occurring at different distances from the galactic centre are likely to show different signatures in their electromagnetic (EM) counterparts, as discussed in the following subsection. {Quick mergers} induced by very strong LK oscillations are naturally produced close to the SMBHB. However, the post-encounter, {slow mergers} created by the secondary SMBH and {ejected mergers} may occur in the outskirts or even outside the host galaxy. 

\begin{figure*}
    \includegraphics[width=\textwidth]{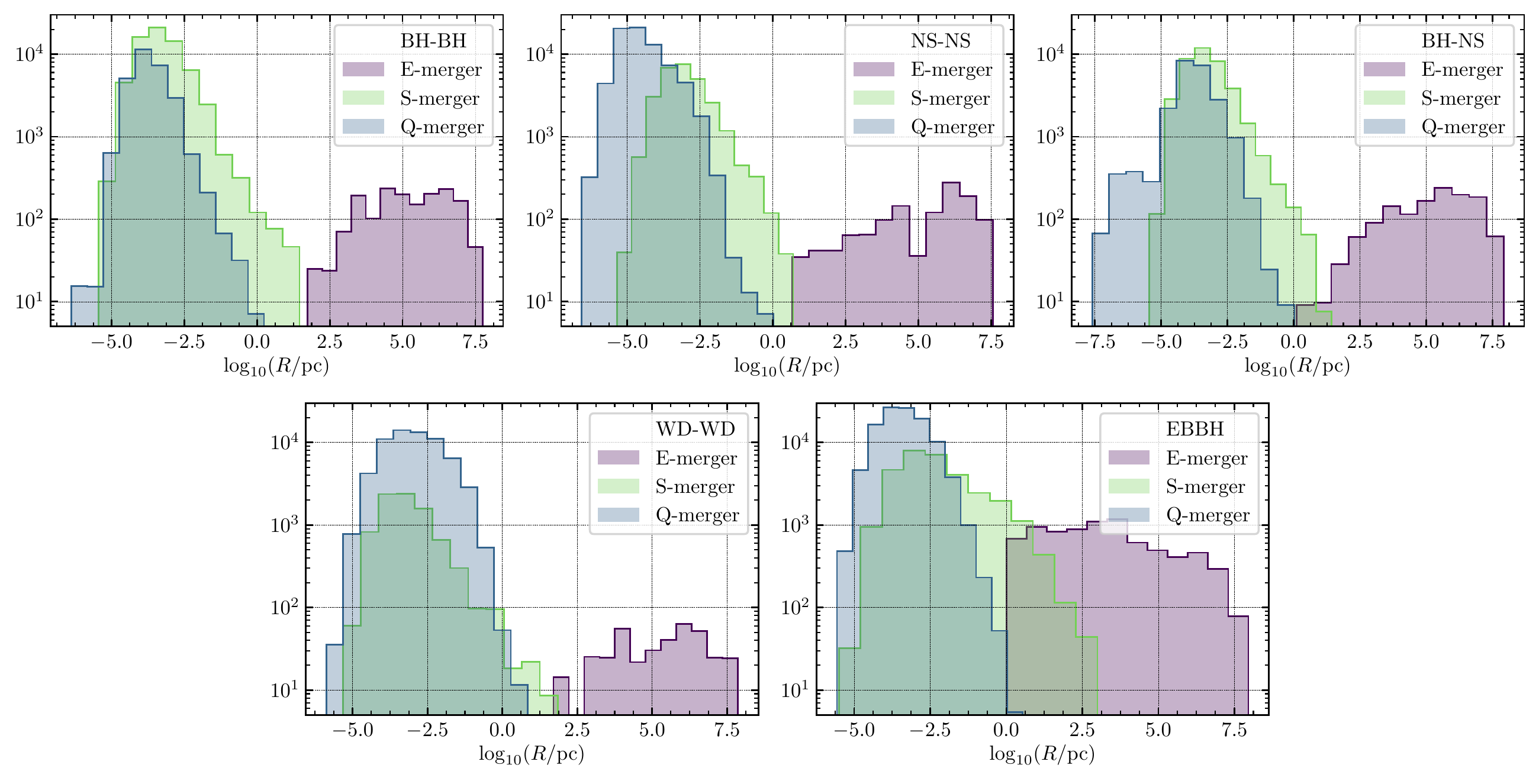}
     \caption{The distribution of the merger distance from the galactic centre $R$ for different types of mergers. \textit{Top Left} panel: BH-BH, \textit{Top middle} panel: NS-NS, \textit{Top right} panel: BH-NS, \textit{Bottom left} panel: WD-WD. \textit{Bottom right} panel: EBBH.}
    \label{fig:r_dist}
\end{figure*}
 
\fig{fig:r_dist} shows the merger distances of quick mergers, {slow} mergers and {ejected} mergers. Independent of the CO species, quick mergers are found within $100~\pc$.  Conversely, all {ejected} mergers will be found at $>10^2\pc$.  As a representative subdivision, we consider $R<10^2\pc$, $10^2\pc<R<10^5\pc$ and $R>10^5\pc$, which correspond to, respectively, the galactic centre, the galactic field and the IGM. The merger rates of different CO binaries in the three regions of parameter space are given in \tab{tab:R}. The parameter $n_{bi}$ quantifies the number of CO binaries per unit mass, and can be determined by combining a specific stellar IMF, with stellar evolution models and empirical estimates for the binary fraction in galactic nuclei.

\begin{table}
\begin{tabular} { | l | l | l | l |}
\hline
  & $R < 10^2\pc$ & $R \in [10^2,10^5]\pc$ & $R > 10^5\pc$ \\
\hline
 \textit{Ejected}:& & &\\

 BH-BH & $1.25 \times 10^{1} n_\mathrm{bh-bh}$ & $7.40 \times 10^{2} n_\mathrm{bh-bh}$& $8.96 \times 10^{2} n_\mathrm{bh-bh}$\\
 NS-NS & $7.65 \times 10^{1} n_\mathrm{ns-ns}$ & $3.87 \times 10^{2} n_\mathrm{ns-ns}$& $6.13 \times 10^{2} n_\mathrm{ns-ns}$\\
 BH-NS & $3.72 \times 10^{1} n_\mathrm{bh-ns}$ & $4.56 \times 10^{2} n_\mathrm{bh-ns}$& $7.20 \times 10^{2} n_\mathrm{bh-ns}$\\
 WD-WD & $3.87 \times 10^{0} n_\mathrm{wd-wd}$ & $1.41 \times 10^{2} n_\mathrm{wd-wd}$& $2.09 \times 10^{2} n_\mathrm{wd-wd}$\\
 Ia SNe & $3.87 \times 10^{0} n_\mathrm{wd-wd}$ & $1.43 \times 10^{1} n_\mathrm{wd-wd}$& $1.83 \times 10^{1} n_\mathrm{wd-wd}$\\
 EBBH & $2.47 \times 10^{3} n_\mathrm{bh-bh}$ & $4.02 \times 10^{3} n_\mathrm{bh-bh}$& $1.38 \times 10^{3} n_\mathrm{bh-bh}$\\

\hline
\textit{Quick}:& & &\\

 BH-BH & $2.83 \times 10^{4} n_\mathrm{bh-bh}$ & $\times$ & $\times$\\
 NS-NS & $7.33 \times 10^{4} n_\mathrm{ns-ns}$ & $\times$ & $\times$\\
 BH-NS & $2.30 \times 10^{4} n_\mathrm{bh-ns}$ & $\times$ & $\times$\\
 WD-WD & $6.45 \times 10^{4} n_\mathrm{wd-wd}$ & $\times$ & $\times$\\
 Ia SNe & $8.72 \times 10^{3} n_\mathrm{wd-wd}$ & $\times$ & $\times$\\
 EBBH & $1.09 \times 10^{5} n_\mathrm{bh-bh}$ & $\times$ & $\times$\\
 
\hline
\textit{Slow}:& & &\\

 BH-BH & $6.63 \times 10^{4} n_\mathrm{bh-bh}$ & $\times$ & $\times$\\
 NS-NS & $2.77 \times 10^{4} n_\mathrm{ns-ns}$ & $\times$ & $\times$\\
 BH-NS & $3.82 \times 10^{4} n_\mathrm{bh-ns}$ & $\times$ & $\times$\\
 WD-WD & $8.42 \times 10^{3} n_\mathrm{wd-wd}$ & $\times$ & $\times$\\
 Ia SNe & $1.30 \times 10^{3} n_\mathrm{wd-wd}$ & $\times$ & $\times$\\
 EBBH & $3.08 \times 10^{4} n_\mathrm{bh-bh}$ & $7.73 \times 10^{1} n_\mathrm{bh-bh}$ & $\times$\\
\hline
\textit{Total}:& & &\\
 BH-BH & $9.47 \times 10^{4} n_\mathrm{bh-bh}$ & $7.39 \times 10^{2} n_\mathrm{bh-bh}$& $8.96 \times 10^{2} n_\mathrm{bh-bh}$\\
 NS-NS & $1.01 \times 10^{5} n_\mathrm{ns-ns}$ & $3.86 \times 10^{2} n_\mathrm{ns-ns}$& $6.12 \times 10^{2} n_\mathrm{ns-ns}$\\
 BH-NS & $6.12 \times 10^{4} n_\mathrm{bh-ns}$ & $4.56 \times 10^{2} n_\mathrm{bh-ns}$& $7.20 \times 10^{2} n_\mathrm{bh-ns}$\\
 WD-WD & $7.29 \times 10^{4} n_\mathrm{wd-wd}$ & $1.41 \times 10^{2} n_\mathrm{wd-wd}$& $2.09 \times 10^{2} n_\mathrm{wd-wd}$\\
 Ia SNe & $1.00 \times 10^{4} n_\mathrm{wd-wd}$ & $1.43 \times 10^{1} n_\mathrm{wd-wd}$& $1.83 \times 10^{1} n_\mathrm{wd-wd}$\\
 EBBH & $1.42 \times 10^{5} n_\mathrm{bh-bh}$ & $4.10 \times 10^{3} n_\mathrm{bh-bh}$& $1.38 \times 10^{3} n_\mathrm{bh-bh}$\\
\hline
\hline
\end{tabular}
\caption{{Model-independent rate of merger events classified by $R$, where $n_\mathrm{wd-wd}$, $n_\mathrm{ns-ns}$, $n_\mathrm{bh-bh}$ and $n_\mathrm{bh-ns}$ are the binary number per unit solar mass in the galactic centre for WD-WD, NS-NS, BH-BH and BH-NS binaries, respectively. The units for all entries are in  $M_\odot\,\yr^{-1}$}.}
\label{tab:R}
\end{table}

The first step in the determination of $n_{bi}$ is to estimate the fraction of single WDs, NSs, and BHs in the galactic nuclei, which requires the evolution of the stellar population from a specific IMF. Since we use the outcome of the {\tt StarTrack} code for our scattering experiments, we follow the same procedure as \citet[][]{2012ApJ...749...91F}. It is generally assumed that the remnant mass is directly determined by the core mass $M_{\rm core}$. Here, we map the IMF to the core mass by using \citep[e.g.,][]{2015MNRAS.451.4086S},
\be
M_{\rm core} = -2 + (B + 2)(g(Z, M_\star; K_1, \delta_1) + g(Z, M_\star; K_2, \delta_2))
\label{eq:mcore}
\ee
where $Z$ is the metallicity, $M_\star$ is the initial stellar mass and
\be
\begin{aligned}
&g(Z, m; x, y) = \frac{0.5}{1+ 10^{(x(Z)-m)y(Z)}}\\
&B=40.98 + 3.415\times 10^4Z - 8.064\times 10^6 Z^2\\
&K_1 = 35.17 + 1.548\times 10^4Z - 3.759\times 10^6 Z^2\\
&k_2 = 20.36 + 1.162\times 10^5Z - 2.276\times 10^7 Z^2\\
&\delta_1 = 2.5\times 10^{-2} -4.346Z +1.34\times 10^3 Z^2\\
&\delta_2 = 1.75\times 10^{-2} + 11.39Z - 2.902\times 10^3 Z^2
\end{aligned}
\ee
for $Z\in[0.1, 0.4] Z_\odot$. The relationship  between the remnant mass and the core mass is
\be
M_{\rm rem} =
\left\{
\begin{aligned}
&0.619 M_\odot, \quad  M_{\rm core} \leq M_{\rm NS,core}^{\rm lower}&\\
&\mathrm{max}\{h(M_{\rm core}, Z), 1.27M_\odot\}, \quad M_{\rm NS,core}^{\rm lower} < M_{\rm core} \leq 5\, M_\odot&\\
&h(M_{\rm core}, Z),\quad 5\, M_\odot < M_{\rm core} \leq 10\, M_\odot&\\
&\mathrm{max}\{h(M_{\rm core}, Z), f(M_{\rm core}, Z)\}, \quad 10\, M_\odot < M_{\rm core}&
\end{aligned}
\right.
\ee
where 
\begin{eqnarray}
h(M_{\rm core}, Z) &=& A_1(Z) + \frac{A_2(Z)- A_1(Z)}{1 + 10 ^{ (L(Z) - M_{\rm core})\eta(Z)  }}\\
f(M_{\rm core}, Z) &=& m(Z)M_{\rm core} + q(Z)\\
m(Z) &=& -6.476\times 10^{2}Z + 1.911\\
q(Z) &=& 2.300\times 10^{3}Z + 11.67\\
A_1(Z) &=& 1.340 - \frac{29.46}{ 1 + (\frac{Z}{1.110\times 10^{-3}})^{2.361}}\\
A_2(Z) &=& 80.22 - 74.73\frac{Z^{0.965}}{2.700\times 10^{-3} + Z^{0.965} }\\
L(Z) &=& 5.683 + \frac{3.533}{ 1 + (\frac{Z}{7.43\times 10^{-3}})^{1.993}}\\
\eta(Z) &=&1.066 - \frac{1.121}{ 1 + (\frac{Z}{2.558\times 10^{-2}})^{0.609}}
\end{eqnarray}
$M_{\rm NS,core}^{\rm lower}$ is the lower mass limit for the NS core, found by substituting $M_\star$ with 
\be
M_{\rm NS}^{\rm lower} = 9M_\odot + 0.9\log_{10}(Z/Z_\odot)
\ee
in equation \eqref{eq:mcore} \citep[e,g,.][]{2012ApJ...749...91F}.

If, we take the Salpeter IMF \citep[e.g.,][]{1955ApJ...121..161S}
\be
\xi(m)dm \propto \bigg(\frac{m}{M_\odot}\bigg)^{-2.35}dm
\ee
with $m_{\rm min} = 0.4\,M_\odot$, $m_{\rm max} = 150\,M_\odot$ and $Z=0.1Z_\odot$ as  typically used in {the initial binary populations from {\tt StarTrack}}, the stellar evolution process yields $1.19$ WDs per unit solar mass and $2.04\times 10^{-2}$ COs per unit solar mass. 

The {\tt StarTrack} code gives the BH-BH, NS-NS and BH-NS number fractions among all CO binaries to be $50\%$, $39.7\%$ and $10.3\%$, respectively. Therefore, if we take the binary number fraction in galactic nuclei to be 0.1 
as a representative value, we find  $n_\mathrm{wd-wd}=5.9\times 10^{-2} \,M_\odot^{-1}$, $n_\mathrm{bh-bh}=1.0\times 10^{-3}\,M_\odot^{-1}$, $n_\mathrm{ns-ns}=8.1\times 10^{-4}\,M_\odot^{-1}$ and $n_\mathrm{bh-ns}=2.1\times 10^{-4}\,M_\odot^{-1}$. The corresponding cosmic merger rates are reported in \tab{tab:rates}. 

\begin{table}
\begin{tabular} { | l | l | l | l |}
\hline
  & $R < 10^2\pc$ & $R \in [10^2,10^5]\pc$ & $R > 10^5\pc$ \\
\hline
 \textit{Ejected}:& & &\\

 BH-BH & $1.25 \times 10^{-2} $ & $7.34 \times 10^{-1} $& $8.96 \times 10^{-1} $\\
 NS-NS & $6.20 \times 10^{-2} $ & $3.13 \times 10^{-1} $& $4.96 \times 10^{-1} $\\
 BH-NS & $7.82 \times 10^{-3} $ & $9.57 \times 10^{-2} $& $1.51 \times 10^{-1} $\\
 WD-WD & $2.29 \times 10^{-1} $ & $8.34 \times 10^{0} $& $1.23 \times 10^{1} $\\
 Ia SNe & $2.29 \times 10^{-1} $ & $8.42 \times 10^{-1} $& $1.08 \times 10^{0} $\\
 EBBH  & $2.47 \times 10^{0} $ & $4.02 \times 10^{0} $& $1.38 \times 10^{0} $\\

\hline
\textit{Quick}:& & &\\

 BH-BH & $2.83 \times 10^{1} $ & $\times$ & $\times$\\
 NS-NS & $5.94 \times 10^{1} $ & $\times$ & $\times$\\
 BH-NS & $4.82 \times 10^{0} $ & $\times$ & $\times$\\
 WD-WD & $3.80 \times 10^{3} $ & $\times$ & $\times$\\
 Ia SNe & $5.14 \times 10^{2} $ & $\times$ & $\times$\\
 EBBH  & $1.09 \times 10^{2} $ & $\times$ & $\times$\\
 
\hline
\textit{Slow}:& & &\\

 BH-BH & $6.63 \times 10^{1} $ & $\times$ & $\times$\\
 NS-NS & $2.25 \times 10^{1} $ & $\times$ & $\times$\\
 BH-NS & $8.02 \times 10^{0} $ & $\times$ & $\times$\\
 WD-WD & $4.97 \times 10^{2} $ & $\times$ & $\times$\\
 Ia SNe & $7.69 \times 10^{1} $ & $\times$ & $\times$\\
 EBBH  & $3.08 \times 10^{1} $ & $7.73 \times 10^{-2} $ & $\times$\\
\hline
\textit{Total}:& & &\\
 BH-BH & $9.47 \times 10^{1} $ & $7.39 \times 10^{-1} $& $8.96 \times 10^{-1} $\\
 NS-NS & $8.19 \times 10^{1} $ & $3.13 \times 10^{-1} $& $4.96 \times 10^{-1} $\\
 BH-NS & $1.29 \times 10^{1} $ & $9.57 \times 10^{-2} $& $1.51 \times 10^{-1} $\\
 WD-WD & $4.30 \times 10^{3} $ & $8.34 \times 10^{0} $& $1.23 \times 10^{1} $\\
 Ia SNe & $5.91 \times 10^{2} $ & $8.42 \times 10^{-1} $& $1.08 \times 10^{0} $\\
 EBBH  & $1.42 \times 10^{2} $ & $4.10 \times 10^{0} $& $1.38 \times 10^{0} $\\
\hline
\hline
\end{tabular}
\caption{Cosmic rate of merger events classified by $R$ with a Salpeter IMF. The units of all entries are in $\yr^{-1}$}
\label{tab:rates}
\end{table}

The same process can also be performed for a log-flat IMF
\be
\xi(m)dm \propto \bigg(\frac{m}{M_\odot}\bigg)^{-1}dm
\ee
which yields $n_\mathrm{wd-wd}=1.7\times 10^{-3}\,M_\odot^{-1}$, $n_\mathrm{bh-bh}=1.9\times 10^{-3}\,M_\odot^{-1}$, $n_\mathrm{ns-ns}=1.3\times 10^{-3}\,M_\odot^{-1}$ and $n_\mathrm{bh-ns}=3.5\times 10^{-4}\,M_\odot^{-1}$. The corresponding cosmic merger rates are listed in \tab{tab:log_flat_rates}.

\begin{table}
\begin{tabular} { | l | l | l | l |}
\hline
  & $R < 10^2\pc$ & $R \in [10^2,10^5]\pc$ & $R > 10^5\pc$ \\
\hline
 \textit{Ejected}:& & &\\

 BH-BH & $2.38 \times 10^{-2} $ & $1.41 \times 10^{0} $& $1.70 \times 10^{0} $\\
 NS-NS & $9.95 \times 10^{-2} $ & $5.03 \times 10^{-1} $& $7.97 \times 10^{-1} $\\
 BH-NS & $1.30 \times 10^{-2} $ & $1.59 \times 10^{-1} $& $2.52 \times 10^{-1} $\\
 WD-WD & $6.59 \times 10^{-3} $ & $2.40 \times 10^{-1} $& $3.11 \times 10^{-1} $\\
 Ia SNe & $6.59 \times 10^{-3} $ & $2.43 \times 10^{-2} $& $3.11 \times 10^{-2} $\\
 EBBH  & $4.70 \times 10^{0} $ & $7.65 \times 10^{0} $& $2.63 \times 10^{0} $\\

\hline
\textit{Quick}:& & &\\

 BH-BH & $5.39 \times 10^{1} $ & $\times$ & $\times$\\
 NS-NS & $9.53 \times 10^{1} $ & $\times$ & $\times$\\
 BH-NS & $8.03 \times 10^{0} $ & $\times$ & $\times$\\
 WD-WD & $1.09 \times 10^{2} $ & $\times$ & $\times$\\
 Ia SNe & $1.48 \times 10^{1} $ & $\times$ & $\times$\\
 EBBH  & $2.07 \times 10^{2} $ & $\times$ & $\times$\\
 
\hline
\textit{Slow}:& & &\\

 BH-BH & $1.26 \times 10^{2} $ & $\times$ & $\times$\\
 NS-NS & $3.60 \times 10^{1} $ & $\times$ & $\times$\\
 BH-NS & $1.34 \times 10^{1} $ & $\times$ & $\times$\\
 WD-WD & $1.43 \times 10^{1} $ & $\times$ & $\times$\\
 Ia SNe & $2.22 \times 10^{0} $ & $\times$ & $\times$\\
 EBBH  & $5.85 \times 10^{1} $ & $1.47 \times 10^{-1} $ & $\times$\\
\hline
\textit{Total}:& & &\\
 BH-BH & $1.80 \times 10^{2} $ & $1.41 \times 10^{0} $& $1.70 \times 10^{0} $\\
 NS-NS & $1.31 \times 10^{2} $ & $5.03 \times 10^{-1} $& $7.97 \times 10^{-1} $\\
 BH-NS & $2.14 \times 10^{1} $ & $1.59 \times 10^{-1} $& $2.52 \times 10^{-1} $\\
 WD-WD & $1.24 \times 10^{2} $ & $2.40 \times 10^{-1} $& $3.56 \times 10^{-1} $\\
 Ia SNe & $1.70 \times 10^{1} $ & $2.43 \times 10^{-2} $& $3.11 \times 10^{-2} $\\
 EBBH  & $2.70 \times 10^{2} $ & $7.79 \times 10^{0} $& $2.63 \times 10^{0} $\\
\hline
\hline
\end{tabular}
\caption{Cosmic rate of merger events classified by $R$ with a log-flat IMF. The units of all entries are $\yr^{-1}$}
\label{tab:log_flat_rates}
\end{table}

\subsection{Electromagnetic counterparts for different types of merger events}

The possibility of drawing astrophysical inferences from the merger
events discussed here largely relies on our ability to identify their merger locations with respect to their galaxy hosts. Studies of host galaxies have a long history in the context of both long and short GRBs, and have been used to probe the nature of their progenitors. The events discussed here span a very large range of distances; however, while {slow} merger events can be produced in a variety  of astrophysical conditions (such as dynamical interactions with a single SMBH), the production of HVBs can only occur in the presence of a SMBHB, and hence are especially interesting, as discussed in this paper.

WD-WD mergers are not observable by LIGO/Virgo in gravitational waves, and hence their EM signatures provide the only available empirical information.  Mergers of two WDs in a binary have been proposed as candidates of Type Ia SNe \citep[the so-called double-degenerate scenario,][]{1984ApJS...54..335I}. 
These transients are typically found to trace the blue light of  galaxies  \citep[e.g.][]{2015yCat..74480732A}, hence  generally occurring within their hosts. The detection of a Type Ia SNe in the IGM would signal an origin from an  HVB,  whether  a  WD-WD  or  a  WD  and  a  companion main  sequence  star \citep[as  in  the  single  degenerate  scenario,][]{1973ApJ...186.1007W}.  

For CO binaries made up of combinations of NSs and BHs, detection by GWs -- when within the LIGO horizon -- constitutes the initial event trigger.  However, localization relies on EM counterparts. Among the three possible types of CO binaries, BH-BH mergers are the least likely to be EM bright. For merger events occurring within an AGN disk, accretion by the local gas may provide some source of accretion at the time of merger \citep{2018arXiv180702859S}. For binaries in the intergalactic medium, on the other hand, special conditions are required, such as, e.g., a fossil disk around one of the two BHs of the binary \citep{Perna2019}, or very strong charges \citep{Zhang2016}.  However, even if such conditions were realized and a relativistic jet was launched, the likelihood of observing it both in $\gamma$-rays as well as at longer wavelengths would be limited by the fact that relatively bright events could only be detected if the jet
points towards the observer. This is because, lacking ejecta material in the close environment of the merger for the jet to interact with, side emission is highly suppressed. Therefore, identification and localization of a BH-BH merger from an HVB in the intergalactic medium may not be feasible.

On the other hand, NS-NS mergers have been confirmed to produce
a relativistic jet and electromagnetic radiation covering a very
broad spectrum, from $\gamma$-rays to the radio band \citep{Abbott2017}.
Even when observed at a sizable angle with respect to the jet axis,
the interaction of the jet with the tidally disrupted material of the NS produces a bright cocoon, which makes these events more easily detectable within the LIGO horizon even for more unfavorable viewing configurations.  In terms of EM radiation, NS-NS mergers from HVBs would be characterized by a much weaker afterglow (i.e. X-rays through radio emission) relative to the prompt $\gamma$-ray emission.  This is because, unlike the $\gamma$-ray radiation, which only depends on the properties of the jet and the ejecta material, the afterglow radiation also depends on the density
$n$ of the medium in which the jet propagates. Lower densities result in dimmer afterglows \citep{Sari1998}. Hence a binary NS merger which is very bright in $\gamma$-rays but very dim at longer wavelengths could signal a merger site in the intergalactic medium, even if not accurately localized due to the lack of a bright afterglow.

Last, the situation for an HVB made up of an NS and a BH is more
complex.  Only binaries with mass ratios $q$ not exceeding $\sim 3-5$ (with the precise value dependent on the equation of state of the NS) are expected to lead to the formation of an accretion disk during the merger, and hence possibly drive a relativistic jet, similar to the NS-NS merger case. On the other hand, large mass ratios will result in the NS being swallowed by the BH without being tidally disrupted. Sources of radiation in this latter case, if at all, would be more similar to those discussed earlier for binary BH mergers.

\section{Conclusions}\label{sec:summary}
In this work we have studied interactions between CO binaries and massive black hole
binaries that may lead to merger events detectable by GW detectors. 
We have performed high-precision N-body simulations in the local reference frame of an SMBHB and combined the scattering outcomes with results from the cosmology simulation Millennium-II to estimate the cosmic rate of different types of CO merger events from dynamical interactions with SMBHBs. Our main conclusions are summarized in the following.

Our simulations show that the interactions between the CO binaries and the SMBHB can produce different types of merger events, which we have dubbed 'quick mergers', '{slow} mergers' and '{ejected} mergers'. Different types of merger events occur at different distances from the galactic centre. Quick mergers with non-negligible eccentricities in the LIGO band can be produced/detected in the innermost region of the SMBHB, in proximity of the primary SMBH. {Slow} mergers, with fully circularized orbits in the LIGO band, can be produced from BSE in the stable region of SMBHBs, or by post-encounters with the secondary SMBH, and can be detected at moderate distances from the galactic centre. The {ejected} mergers that are produced from the hypervelocity binaries could potentially be detected in the galactic halo or in the IGM. Different types of merger events can be precisely classified by the stability map of the SMBHB introduced in Section  \ref{sec:stability} and might be distinguished observationally by measuring the distribution of merger distances to the galactic centre from EM counterparts, if available.

Close encounters between CO binaries and SMBHBs can create hypervelocity binaries that could escape the host galaxies and eventually merge in the intergalactic medium. Such a high velocity cannot be produced by interactions in globular clusters nor multiple stellar systems; thus an off-centre merger event might indicate the existence of a nearby SMBHB, especially a high mass ratio SMBHB formed from a prior major galaxy-galaxy merger, which would kick out any merged CO binaries occurring in its orbital plane.

Assuming a binary fraction $n_\bi$ of $10\%$ and a Salpeter IMF, we estimate the merger rate of WD-WD binaries to be $4.32\times 10^{3}\yr^{-1}$, of which $12.3\yr^{-1}$ occur in the inter-galactic medium, $8.34\yr^{-1}$ occur in the DM halo, and $4.30\times 10^{3}\yr^{-1}$ occur in the galactic nucleus. The merger rate of BH-BH binaries is $96.3\yr^{-1}$, of which $0.896\yr^{-1}$ occur in the inter-galactic medium, $0.740\yr^{-1}$ occur in the DM halo, and $94.7\yr^{-1}$ occur in the galactic nucleus. The merger rate of NS-NS binaries is $82.7\yr^{-1}$, of which $0.496\yr^{-1}$ occur in the inter-galactic medium, $0.313\yr^{-1}$ occur in the DM halo, and $81.9\yr^{-1}$ occur in the galactic nucleus. The merger rate of EBBHs is $148\yr^{-1}$, of which $1.08\yr^{-1}$ occur in the inter-galactic medium, $0.842\yr^{-1}$ occur in the DM halo, and $142\yr^{-1}$ occur in the galactic nucleus. Finally, the merger rate of BH-NS binaries is $13.1\yr^{-1}$ of which $0.151\yr^{-1}$ occur in the inter-galactic medium, $0.0957\yr^{-1}$ occur in DM halos and $12.9\yr^{-1}$ occur in galactic nuclei.

 Finally, we caution that our scattering experiments can only describe the cusp erosion phase, in which the SMBHB ejects stars in the bound cusp surrounding it as it first becomes bound. Further ejection of unbound CO binaries in the later stages of the SMBHB evolution might contribute to the cosmic rate of fast, slow and ejected CO binary mergers. We defer the study of this later phase to future work.

\section*{Acknowledgements}
RP and NL acknowledge support by NSF award AST-1616157.
The Center for Computational Astro-physics at the Flatiron Institute is supported by the Simons Foundation.



\appendix


\bsp
\label{lastpage}

\bibliography{refs} 

\end{document}